\documentclass[10pt,conference]{IEEEtran}
\IEEEoverridecommandlockouts

\usepackage{tcolorbox}
\usepackage{soul}
\usepackage[noend]{algorithm2e}
\usepackage{stfloats}
\usepackage{caption}
\usepackage{subcaption}
\usepackage{pifont}
\usepackage{listings}
\usepackage{makecell}
\usepackage{colortbl}
\usepackage{diagbox}
\usepackage{multirow}
\usepackage{adjustbox}
\usepackage[hang,flushmargin]{footmisc} 

\usepackage[noadjust]{cite}
\usepackage{amsmath,amssymb,amsfonts}
\usepackage{graphicx}
\usepackage{textcomp}
\usepackage{xcolor}
\usepackage[colorlinks=true,linkcolor=black,anchorcolor=black,citecolor=black,filecolor=black,menucolor=black,runcolor=black,urlcolor=black]{hyperref}
\usepackage[moderate,tracking=normal,mathdisplays=normal]{savetrees}

\definecolor{dkgreen}{rgb}{0,0.6,0}
\definecolor{gray}{rgb}{0.5,0.5,0.5}
\definecolor{mauve}{rgb}{0.58,0,0.82}
\definecolor{gray}{rgb}{0.4,0.4,0.4}
\definecolor{darkblue}{rgb}{0.0,0.0,0.6}
\definecolor{lightblue}{rgb}{0.0,0.0,0.9}
\definecolor{cyan}{rgb}{0.0,0.6,0.6}
\definecolor{darkred}{rgb}{0.6,0.0,0.0}

\lstdefinelanguage{XML}
{
    morestring=[s][\color{mauve}]{"}{"},
    morestring=[s][\color{black}]{>}{<},
    morecomment=[s]{<?}{?>},
    morecomment=[s][\color{dkgreen}]{<!--}{-->},
    stringstyle=\color{black},
    identifierstyle=\color{lightblue},
    keywordstyle=\color{red},
    morekeywords={xmlns,xsi,noNamespaceSchemaLocation,type,id,x,y,source,target,version,tool,transRef,roleRef,objective,eventually,name}
}

\definecolor{delim}{RGB}{20,105,176}
\definecolor{numb}{RGB}{106, 109, 32}
\definecolor{string}{rgb}{0.64,0.08,0.08}

\lstdefinelanguage{json}{
    morestring=[b]",
    stringstyle=\color{string},
    literate=
     *{0}{{{\color{numb}0}}}{1}
      {1}{{{\color{numb}1}}}{1}
      {2}{{{\color{numb}2}}}{1}
      {3}{{{\color{numb}3}}}{1}
      {4}{{{\color{numb}4}}}{1}
      {5}{{{\color{numb}5}}}{1}
      {6}{{{\color{numb}6}}}{1}
      {7}{{{\color{numb}7}}}{1}
      {8}{{{\color{numb}8}}}{1}
      {9}{{{\color{numb}9}}}{1}
      {\{}{{{\color{delim}{\{}}}}{1}
      {\}}{{{\color{delim}{\}}}}}{1}
      {[}{{{\color{delim}{[}}}}{1}
      {]}{{{\color{delim}{]}}}}{1},
}

\definecolor{mBasic}{RGB}{248,248,242}       
\definecolor{mKeyword}{RGB}{0,0,255}          
\definecolor{mString}{RGB}{160,32,240}        
\definecolor{mComment}{RGB}{34,139,34}        
\definecolor{mBackground}{RGB}{245,245,245}   
\definecolor{mNumber}{RGB}{128,128,128}       

\lstdefinelanguage{Matlab}{
    keywordstyle={\color{mKeyword}},     
    stringstyle={\color{mString}},       
    commentstyle={\color{mComment}},     
    morecomment=[l][\color{mComment}]{\%},
    keywords={break,case,catch,classdef,continue,else,elseif,end,for,
    function,global,if,otherwise,parfor,persistent,return,spmd,switch,try,while}
}  

\SetKwComment{Comment}{/* }{ */}
\RestyleAlgo{ruled}

\usepackage{etoolbox}
\providetoggle{beginFlag}
\settoggle{beginFlag}{true}

\SetAlgoSkip{SkipBeforeAndAfter}

\newcommand{\name}{SA-MCAS}
\newcommand{\mcasold}{MCAS$_{old}$}
\newcommand{\mcasnew}{MCAS$_{new}$}

\makeatletter
\renewcommand{\@IEEEsectpunct}{~}
\makeatother

\AtBeginDocument{%
\abovedisplayskip=1pt plus 3pt
\belowdisplayskip=1pt plus 3pt
\abovedisplayshortskip=1pt plus 3pt
\belowdisplayshortskip=1pt plus 3pt
}

\begin{document}

\date{}

\title{
    Analysis and Prevention of MCAS-Induced Crashes
\thanks{{\bf Acknowledgements}: The work in this paper was supported in part by ONR under Grant No. N00014-22-1-2622.}
}

\author{
\IEEEauthorblockN{Noah T. Curran, Thomas W. Kennings, Kang G. Shin}
\IEEEauthorblockA{
    \textit{Computer Science and Engineering} \\
    \textit{University of Michigan--Ann Arbor} \\
    {\{ntcurran, kennings, kgshin\}@umich.edu} \\
}
}


\maketitle

\begin{abstract}
Semi-autonomous (SA) systems face the challenge of
determining which source to prioritize for control,
whether it's from the human operator or the
autonomous controller, especially when they conflict
with each other.
While one may design an SA system 
to default to accepting control from one or the other, 
such design choices can have catastrophic
consequences in safety-critical settings.
For instance, the sensors an autonomous controller
relies upon may provide incorrect information about
the environment due to tampering or natural fault.
On the other hand, the human operator may also provide
erroneous input.

To better understand the consequences and resolution
of this safety-critical design choice, we investigate
a specific application of an SA system that failed due to 
a static assignment of control authority: the
well-publicized Boeing 737-MAX Maneuvering 
Characteristics Augmentation System (MCAS) that caused the 
crashes of Lion Air Flight 610
and Ethiopian Airlines Flight 302.
First, using a representative
simulation, we analyze  and demonstrate the ease
by which the original MCAS design could fail.
Our analysis reveals the most robust public analysis
of aircraft recoverability under MCAS faults, offering 
bounds for those scenarios beyond the original crashes.
We also analyze Boeing's updated MCAS and 
show how it falls short of its intended goals
and continues to rely upon on a fault-prone
static assignment of control priority.
Using these insights, we present Semi-Autonomous MCAS 
(\name), a new MCAS that {\it both} meets the intended 
goals of MCAS {\it and} avoids the failure cases that 
plague both MCAS designs. We demonstrate \name's 
ability to make safer and timely control decisions of 
the aircraft, even when the human and autonomous operators 
provide conflicting control inputs.
\end{abstract}


\section{Introduction}
\label{sec:intro}
Semi-autonomous (SA) systems---those that take both autonomous
and manual inputs to control their actions---are ubiquitous 
in the modern world, presenting applications in factories, 
hospitals, transportation, and more. 
Often, the purpose of these systems is to improve the 
safety and efficiency of tasks that take substantial 
manual effort. 
Airplanes, for example, use SA control to maintain safe 
flight while pilots perform other tasks. As a consequence of
SA systems' close coupling with safety-critical applications,
there is a complicated trade-off between trusting human and 
autonomous inputs. SA functionality is often included in 
a system because humans are prone to making mistakes, but 
autonomous systems are also imperfect.
These autonomous controllers serve as a safety-critical
component of embedded systems, and as such, their design
should be closely scrutinized.

In order to handle these cases where there is conflict
in control, it is necessary for an SA
embedded system to 
incorporate a routine for resolving the conflict.
Oftentimes, such a routine is hard-coded to always ``trust" 
the input from one entity over the other. For instance, 
an autonomous system in the Boeing 737-MAX, the 
{\em Maneuvering Characteristics Augmentation System} 
(MCAS), an embedded system controller,
was originally given static priority to manually
override the pilot's control of the aircraft during 
aircraft stall events.
While this decision was motivated by a lack of 
trust in pilot control during these safety-critical 
stall events, it had dire consequences. 
In 2019, MCAS mistakenly identified a stall event
due to faulty sensor data, which ultimately
caused the crash of  two Boeing 737-MAX aircraft:
Lion Air Flight 610 (JT610)
and Ethiopian Airlines Flight 302 (ET302).
These crashes prompted regulators to mandate Boeing
redesign the MCAS before the 737-MAX could fly
again~\cite{faa:19:grounding}.
Among the numerous changes Boeing made, one removed
the static control priority from MCAS and gave it
to the pilots~\cite{boeing:19:mcas_updates}.

In practical settings, the consequences from static
assignment of control priority in
embedded systems is not well documented.
Leveraging the circumstances leading to Boeing creating
two versions of MCAS that have opposing static priority
controllers, we study the differences between their
control failures under various fault types 
(\autoref{sec:background:motivation}).
From our analysis, we conclude that the redesign
of MCAS takes an incorrect approach. 
We find that giving one entity the
capability to override control of a safety-critical 
application creates a single point-of-failure,
even in the case of the current MCAS version.
Rather than defaulting the control to one entity,
we argue for a dynamic control conflict arbiter
that chooses which entity to allow control based on the
aircraft's situation.
This removes the single point-of-failure from MCAS,
making the aircraft more tolerant of erroneous input
from either autonomous or manual control.

Following this embedded system controller
design philosophy, we propose a version
of MCAS with a dynamic control arbiter, which we call
Semi-Autonomous MCAS (\name).
Unlike the prior implementations of MCAS, \name{} is 
capable of providing safer control of the aircraft's pitch in 
the presence of erroneous input from {\em both}
autonomous {\em or} manual control.
Through our investigation, we demonstrate that \name{}
can select which operator to control the
aircraft in the presence of erroneous sensor
readings or erroneous pilot control.
We test the robustness of \name{} under a dataset
of representative flight scenarios under which MCAS
may activate. Under these flight scenarios, we subject
\name{} to numerous data, control, and timing errors,
summarized in \autoref{sec:threat_model}.

{\color{red}


}

Previously, there have been a few publicized analyses and 
reports of MCAS~\cite{gates:20:st,lemme:21:mid_value}.
Mainly, these were conducted by aeronautic enthusiasts 
and reporters who wished to demonstrate {\em how} the 
crash occurred. They sparsely discuss the choices made in 
the design of the embedded control logic without considering
or mentioning alternative designs. In contrast, we present
the first in-depth analysis of the control logic of MCAS,
with the goal of finding a design alternative 
with better overall safety.

Prior work on the resolution of control conflict of SA systems
is scarce. In the context of cars, there is some prior
work on conflict resolution between autonomous
controllers and drivers~\cite{chen:22:thesis}. Alongside this
line of work, there is research in sensor anomaly
detection~\cite{xue:22:usenixsec,fei:19:tvt,curran:23:cns},
estimation~\cite{ganesan:17:sae,curran:23:cns},
and fault injection~\cite{checkoway:10:oakland,miller:15:blackhat}.
While this work is somewhat relevant to our research, we focus 
on arbitration of conflicting autonomous and manual control.

This  paper makes the following novel contributions:
\begin{enumerate}
    \item We build a MATLAB/Simulink template\footnote{Available on
        GitHub: \url{https://github.com/noah-curran/SA-MCAS}.\label{fn:github}}
        for simulating control input
        for aircraft modeled in JSBSim~\cite{berndt:04:aiaa}.
        We provide the building blocks for easily creating and 
        evaluating new aircraft control systems.
        (\autoref{sec:sim})
    \item We model timing and control constraints on
        the aircraft, determining the parameter boundaries
        for the recoverability of safe control of the aircraft.
        We conduct this for Boeing's original (\mcasold),
        new (\mcasnew), and our (\name) versions of MCAS. 
        Our analysis tweaks
        the MCAS response time, the MCAS duration,
        the fixed time interval between MCAS
        events, and the pilot's reaction delay to MCAS. 
        (\autoref{sec:threat_model} and \autoref{sec:threats})
    \item For the first time, we model a
        comprehensive failure analysis for both \mcasold{} 
        and \mcasnew{} in the Boeing 737-MAX.
        This model incorporates incorrect sensor data that
        MCAS relies upon and dangerous flight control that
        leads to stalls. 
        Our analysis uncovers previously unseen
        flight scenarios resulting in aircraft crashes due to
        erroneous control input, in addition to the
        scenarios that occurred in the original Boeing
        737-MAX crashes.
        (\autoref{sec:threat_model} and \autoref{sec:threats})
    \item We propose \name, which makes control  
        decisions that account for erroneous
        inputs from the pilot or autonomous system.
        \name{} is shown to improve the 
        state-of-the-art, loosening the constraints for
        recoverability of flights to safe control in
        comparison to \mcasold{} and \mcasnew{}.
        (\autoref{sec:prevention})
\end{enumerate}





\section{Background \& Motivation}
\label{sec:background}
For better comprehension of the problem, we provide the necessary
knowledge for understanding the Boeing 737-MAX crashes.
We start with a basic explanation of longitudinal flight
dynamics. Such information is essential for understanding
why Boeing chose to include MCAS in the 737-MAX and for
understanding how MCAS impacts the aircraft's control.
We then discuss the brief history of the MCAS
implementation and how its design flaws caused the
crashes of JT610 and ET302. Since these
crashes, both the FAA and Boeing have introduced
new requirements for MCAS to avoid future
accidents. We analyze these new requirements 
to motivate our investigation.

\subsection{System Components}
\label{sec:background:system}
\begin{figure}
    \centering
    \includegraphics[width=0.8\columnwidth]{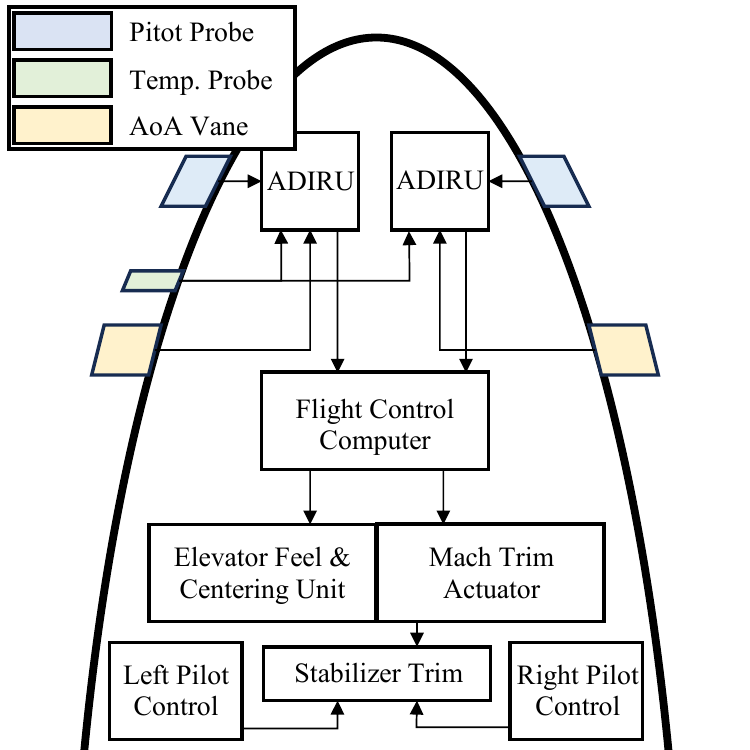}
    \caption{\small The components of the Boeing 
    737-MAX aircraft pitch control stability system.
    This was adapted from those found 
    in~\cite{boeing:operations}.}
    \label{fig:system}
    \vspace{-0.3in}
\end{figure}

The components of the aircraft system that are of concern to
this work are summarized in \autoref{fig:system}.
The {\em Air Data Inertial Reference Unit} (ADIRU) is at the
core of the flight control system. It is information supplied 
from various sensors, which enables the calculation of the
airspeed, angle-of-attack (AoA), altitude, position, and other
inertial and environmental information.
As seen in \autoref{fig:system}, there is a redundant ADIRU
available in the Boeing 737-MAX.

The information from the ADIRU is passed along to the
flight control computer, which is responsible for
actuating the aircraft control surfaces. In the case of
pitch control, it uses the Mach information from the ADIRU
to trim the {\em horizontal stabilizer} (HS). The flight
dynamics and control that govern this process are described
in the following subsection. Moreover, the pilots can
manually trim the HS and compete for control
of the aircraft with the flight control computer.

\subsection{Longitudinal Flight Dynamics \& Control}
\label{sec:background:dynamics}
The longitudinal flight dynamics involve moments about the aircraft's $Y$-axis.
The longitudinal equations\footnote{See~\cite{how:13:mitocw}
for the full longitudinal equations.}
describe the physics of the angular velocity of the pitch 
($\dot{\theta}$), the angular acceleration of the pitch ($\dot{q}$), 
the acceleration ($\dot{U}$) in the $X$-direction, and the
acceleration ($\dot{W}$) in the $Z$-direction. These
equations can be reduced to state-space form as
\begin{equation}
    \begin{bmatrix}
    \dot{U} & \dot{W} & \dot{q} & \dot{\theta}
    \end{bmatrix}^\intercal = A\begin{bmatrix}
    U & W & q & \theta
    \end{bmatrix}^\intercal + Bu,
\end{equation}
where $A$ is the state matrix, $B$ is the input matrix,
and $u = \begin{bmatrix} \delta_e & \delta_p
\end{bmatrix}^\intercal$ is the control vector. This is a 
linear system that is controlled by the deflection of the 
elevator ($\delta_e$) and the change in the thrust ($\delta_p$).
To examine control design, we can approximate the
longitudinal dynamics using simple models. For our
purposes, we can focus on the short period 
approximation~\cite{how:13:mitocw}.
For the short period approximation, the response from
$\theta$ and $w$ are in the same phase, and the response
from $u$ and $q$ are very small. The resulting equation is
\begin{equation}
    \dot{x}_{sp} = \begin{bmatrix}
    \dot{w} & \dot{q}
    \end{bmatrix}^\intercal = A_{sp}x_{sp} + B_{sp}\delta_e.
    \label{eqn:shrt_prd}
\end{equation}
Here, we assume that the thrust remains constant in steady
flight. The control input of interest is the deflection of
the elevator, as this is what the pilot uses to directly
impact the pitch of the aircraft. The elevator is attached
to the HS, which is what
MCAS controls in order to automatically trim the
aircraft's pitch.

\subsection{MCAS} 
\label{sec:background:mcas}

\subsubsection{Why Was MCAS Necessary?}

During the design of an aircraft, regulatory bodies will
provide a type certificate once it is deemed safe for
flying in the air. If a newly designed aircraft is the
same type as a previously certified aircraft, the regulation
is expedited since certain preliminary prototypes are
unnecessary. Instead, just the parts of the aircraft that
are changed need to be tested. An additional benefit of
this practice is the reduction in the amount of training
the pilots of the previous aircraft require in order to
operate the new aircraft. The type certificate is amended
in order to include the updates made.

During the design of the 737-MAX line of aircraft, Boeing
sought to certify it as a 737 variant to take advantage of
this certification amendment process. One such change was 
made to the engines, in which the 737-MAX moved from the
CFM56 engine to the LEAP-1B engine, which is larger and
placed farther forward on the wings of the aircraft.
Testing revealed that when the 737-MAX encountered
high-pitch scenarios at low airspeeds,
the weight distribution of the new engines 
would push the nose of the 737-MAX
further upward and cause the aircraft to gain a
high Angle-of-Attack (AoA) and subsequently stall.\footnote{A
stall is a flight event where there is not enough lift
under the wing of the aircraft, causing the aircraft to lose
altitude. These normally occur when the AoA is 
$\gtrsim$17$^\circ$.} To address this issue, Boeing
would either need to (A) redesign the entire aircraft
type to accommodate the engines and follow a long and
expensive type certification process, or (B) use some
flight control mechanism to counter the problematic
stall behavior. Since the former option would compromise
Boeing's goals to get the 737-MAX on the market quickly
and reduce the costs of pilot training, they
provided the aircraft with \mcasold, a flight stabilization
program that automatically pitches the aircraft down to
prevent a stall during high-AoA maneuvering.

\subsubsection{How Does MCAS Work?}

In order to determine whether a stall event is about to
occur, MCAS observes the AoA of the aircraft through a
sensor. The AoA sensor is a swept vane that is aerodynamically
aligned with the aircraft in order to measure the angle of the
airflow passing the wing. While the 737-MAX
is equipped with two AoA sensors (one on both sides of the
nose of the plane), \mcasold{} used just one of
the sensors. In response to a high AoA (defined
as $\sim$17$^\circ$),
\mcasold{} provides a nose-down
control input to the HS to avoid a stall. At low
speeds, this nose-down deflection is 2.5$^\circ$ and at
high speeds it is 0.6$^\circ$~\cite{gates:20:st}.
During the high-speed stall events, \mcasold{} checks for a high
$g$-force in addition to the AoA. The $g$-force check is
omitted during low-speed flight.

\subsubsection{What are the Functional Requirements of MCAS?}
For the functional safety of any part aboard an aircraft,
the FAA follows ARP4761 and ARP4754 to provide an assurance
level for the design~\cite{ARP4761,ARP4754}.
In the case of MCAS, the FAA designated it as a ``hazardous 
failure" system, which should have a probability of occurring
at $<10^{-7}$ per flight hour.
In this case, ``failure has a large negative 
impact on safety or performance, or reduces the ability of the 
crew to operate the aircraft due to physical distress or a 
higher workload, or causes serious or fatal injuries among the 
passengers." These safety requirements assume an undistracted
pilot can respond to an issue within 3 
seconds~\cite{gates:20:st}.

\subsubsection{How Did MCAS Cause Deadly Accidents?}
The issues with \mcasold{} occurred due to design choices of its
activation during low-speed stall events. Because there was
no $g$-force check and only one AoA sensor was checked, a
single-point-of-failure was present.
The $g$-force check is primarily to determine whether 
pilots need assistance during high-speed events,
which cause the pitch control column
to become too heavy and cumbersome to control.
Boeing's initial disclosure of MCAS to the FAA accounted
for its necessity to activate during high-speed stall
events. The failure analysis from Boeing's disclosure
demonstrated that \mcasold{} was less intrusive to the 
flight controls (and deemed a low risk) due to the 
improbability of its failure and due to the redundancy 
the $g$-force check provided during the high-speed 
stall events.
While the issues experienced during high-speed stall 
events were a non-issue during low-speed stall events,
a high $g$-force could still be useful as a source of
redundancy to detect the stall event.
The original authority \mcasold{} was granted to provide nose-down
deflection to the HS was limited to 0.6$^\circ$.
However, after discovering issues during low-speed tests, Boeing
provided \mcasold{} additional authority.
In short, during low speeds the pitch control surface of an
aircraft requires more deflection in order to yield the same
response as during high speeds, so Boeing increased the nose-down
deflection amount to 2.5$^\circ$ for low-speed events.
After making this substantial change to \mcasold{}, Boeing failed
to notify the FAA and made no mention of \mcasold{} in the 737-MAX's
pilot manuals.

The increased likelihood of a dangerous event occurring compounded
with pilots inadequately prepared to handle an
erroneously engaged \mcasold{} created a recipe for disaster.
Two deadly crashes followed: ET302 and JT610.
During these flights, the AoA sensor delivered faulty
readings that made \mcasold{} believe the
airplane's AoA was too high. 
Consequently, the nose of the aircraft was
pushed down by \mcasold{}. 
To counteract \mcasold{}, the pilot manually
trimmed the HS and pulled
back on the column to actuate the elevator to undo the nose-down
deflection. \mcasold{} again displaced the HS due to the sensor's
incorrect readings, entering a state known
as a ``runaway stabilizer".
After back-and-forth between \mcasold{} and the pilot, the HS was
eventually displaced so much that elevator deflection could not
counter the effects of the much larger HS.
Also, due to aerodynamic factors, the manual HS hand-crank
available in the cockpit eventually would not budge.
In both catastrophic cases, the aircraft entered a steep
nosedive and crashed. The two crashes killed all 346 people
onboard and resulted in the grounding of all Boeing 737-MAX
aircraft globally. While skilled pilots were sometimes capable of
landing aircraft that \mcasold{} negatively
impacted, these instances 
were not reported to any regulatory agencies until after 
the deadly crashes~\cite{jt610:18:report}.

\subsubsection{How Did MCAS Change After the Crashes?}

In response to these crashes, Boeing proposed a redesigned
\mcasnew{} with several changes~\cite{boeing:19:mcas_updates}.
First, MCAS will now check both the left and right
AoA sensors. If the AoA sensors exceed $17^\circ$ when
the flaps are not up or if they disagree with one
another more than $>5.5^\circ$, \mcasnew{} will not activate.
Additionally, Boeing introduced Mid-Value Select (MVS)
to pick an AoA value when they disagree within the acceptable
range~\cite{lemme:21:mid_value}.
\mcasnew{} will store the previously selected AoA
value and during the next iteration it will pick the median
between the stored, left, and right AoA values.
Second, \mcasnew{} will only activate once per sensed
event rather than an unconstrained number of times, preventing
a runaway stabilizer. Lastly, when \mcasnew{} does engage, pilots
can now override it and perform manual flight at any time
since it will not provide more input on the HS than the
pilot can put on the elevator. The final revision approved
by the FAA included an additional requirement to the flight
control computer, requiring an integrity monitor in order
to stop erroneously generated trim commands from
\mcasnew~\cite{faa:20:faa_mcas_fix}.



\subsection{Analysis of the Revised MCAS Requirements}
\label{sec:background:motivation}
Boeing's revised requirements for \mcasnew{} made
a major pivot in control authority.
\mcasold{} was originally designed with absolute
authority; in fact, there was not even a switch for cutting off
its control of the HS in the event of a runaway stabilizer.
However, in \mcasnew, there is a clear lack of trust for
autonomous control reversing control authority in
favor of the pilot. This change is notable, as it is contrary
to Boeing's original goals of providing mechanisms to emulate
the feel of the 737-NG and reducing the amount of training
required for the pilot. Now, the pilot must learn how to safely
counter the MCAS in the event of additional problems occurring.


These additional problems are not completely unimaginable.
While the choice to use an agreement between the left
and right AoA sensors to validate their use
is an improvement over the single-point-of-failure,
the AoA sensors are subject to the same environmental factors
and hence the two sensors may possibly agree with each
other on an incorrect value. This differs from
the case of MCAS activation at high-speed, which is contingent
on a high $g$-force value as well. Rather than seeking a source
of redundancy within an entirely separate system, Boeing chose
to consider just the unused AoA sensor.

Such scenarios where both the autonomous entity and a human
compete for control of a vehicle is called a {\em semi-autonomous}
(SA) system. Defaulting control authority to one entity in the case
of disagreement is a common trend in SA system design.
While Boeing designed \mcasold{} and \mcasnew{} with this default
behavior, one can see cases where the pilot is more trustworthy
and others where the MCAS is: there are instances in flight where
either the pilot or the autonomous system could be incorrect.
On one hand, sensor failures have and will continue to occur, and
on the other hand, pilots may not respond quickly or
correctly enough during chaotic flight scenarios.
Thus, we argue that neither the original design nor the
redesign of MCAS is the right choice.

\begin{figure}
    \centering
    \begin{subfigure}[b]{0.48\columnwidth}
        \centering
        \includegraphics[width=\textwidth]{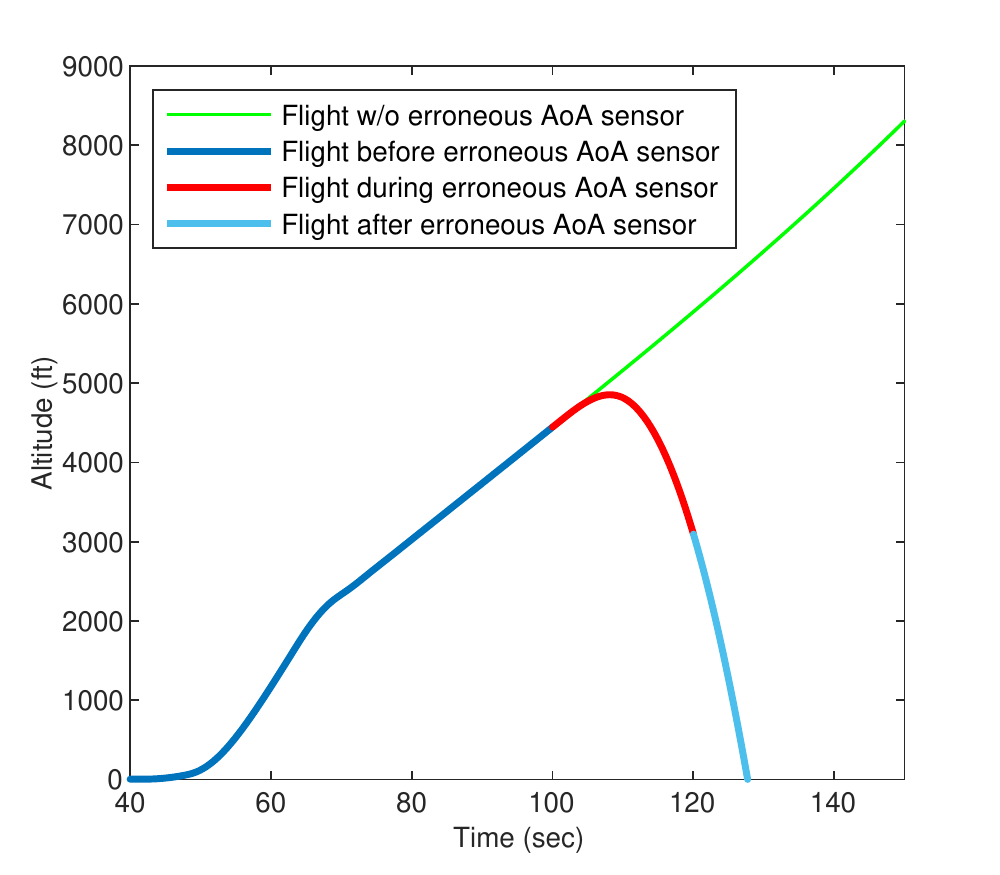}
        \caption{\footnotesize Erroneous AoA values for 20 seconds 
        with \mcasold.}
        \label{fig:prestudy:pre_crash__aoa_error}
    \end{subfigure}
    \hfill
    \begin{subfigure}[b]{0.48\columnwidth}
        \centering
        \includegraphics[width=\textwidth]{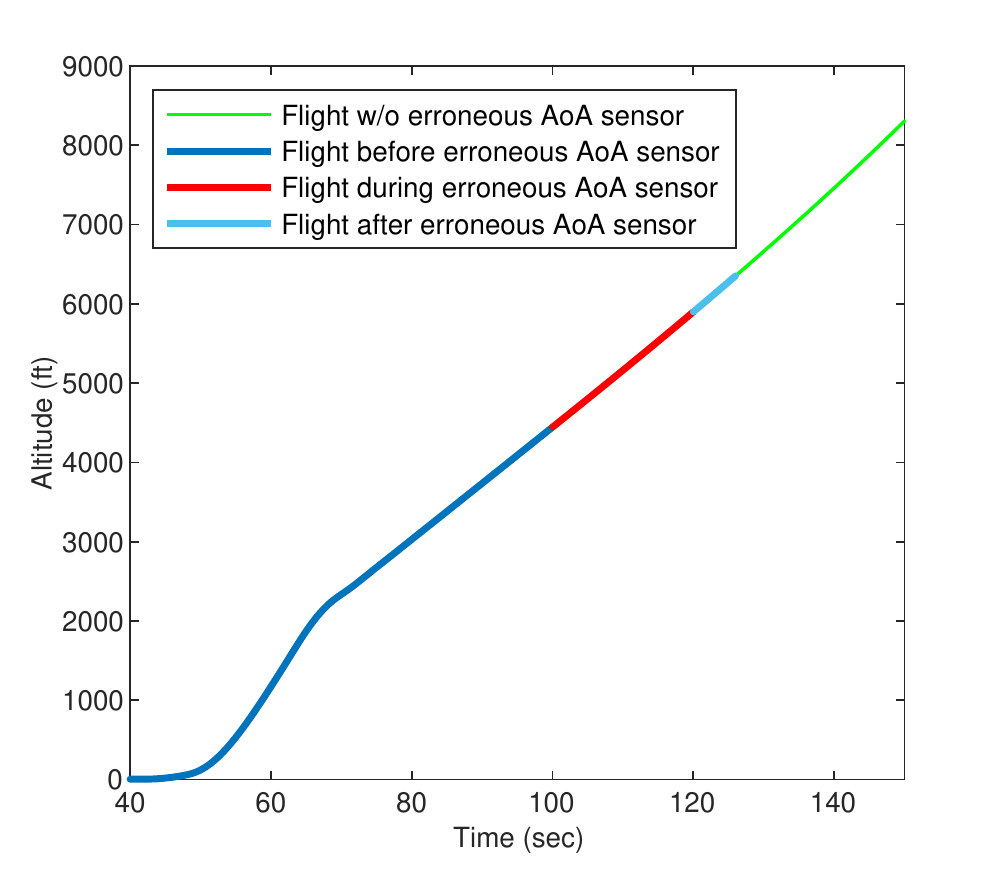}
        \caption{\footnotesize Erroneous AoA values for 20 seconds 
        with \mcasnew.}
        \label{fig:prestudy:post_crash_mcas_aoa_error}
    \end{subfigure}
    
    \begin{subfigure}[b]{0.48\columnwidth}
        \centering
        \includegraphics[width=\textwidth]{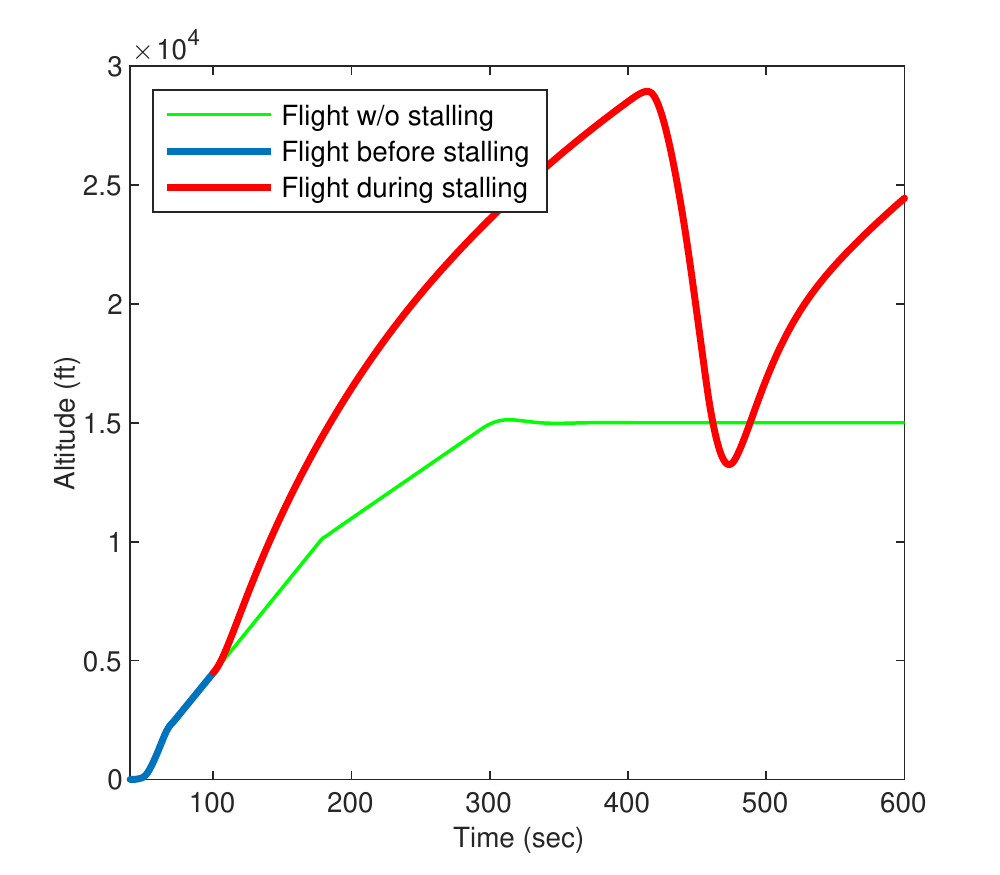}
        \caption{\footnotesize Pilot stalling aircraft with high pitch and \mcasold.}
        \label{fig:prestudy:pre_crash_mcas_pilot_error}
    \end{subfigure}
    \hfill
    \begin{subfigure}[b]{0.48\columnwidth}
        \centering
        \includegraphics[width=\textwidth]{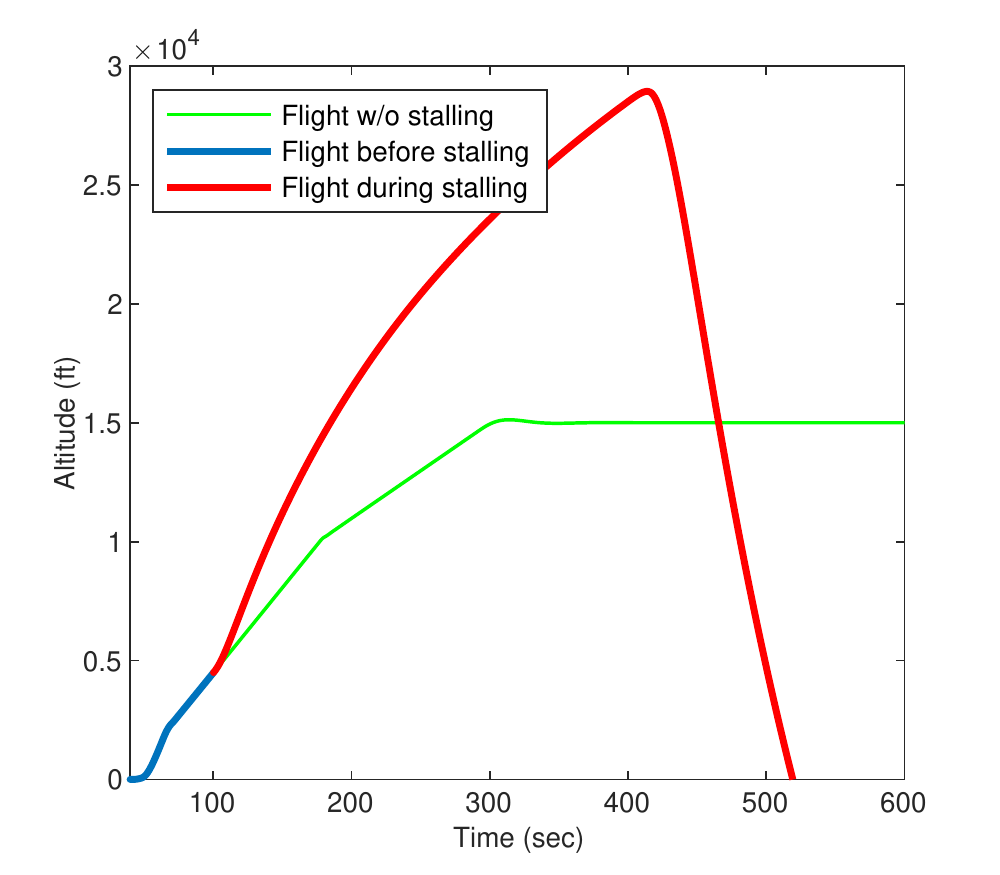}
        \caption{\footnotesize Pilot stalling aircraft with high pitch and \mcasnew.}
        \label{fig:prestudy:post_crash_mcas_pilot_error}
    \end{subfigure}
    \caption{\small Simulation of the fault-tolerance of \mcasold{} and \mcasnew. For each,
    we simulated 737-MAX takeoff using our custom toolkit 
    (\autoref{sec:sim}) built on top of
    JSBSim~\cite{berndt:04:aiaa}.
    }
    \label{fig:prestudy}
    \vspace{-0.3in}
\end{figure}

To back this claim, we
reconstruct the behavior of \mcasold{} and \mcasnew{}
using information {\it The Seattle Times} \cite{gates:20:st}
and the FAA~\cite{faa:20:faa_mcas_fix} reported to the public
and conduct a preliminary analysis (\autoref{fig:prestudy}).
Using the open-source flight dynamic simulator
JSBSim~\cite{berndt:04:aiaa}, we built a toolkit for 
running MCAS experiments
(see \autoref{sec:sim} for more details of how it works).
The preliminary analysis is our first step to investigate
the safety of the \mcasnew. The simple experiment runs a
takeoff maneuver with either an injection of an erroneous
AoA sensor at $t=100$s or a pilot beginning to stall
the aircraft with a high pitch at $t=100$s.
While \mcasnew{} mends the original single-point-of-failure
issue, it introduces a new hazard related to pilot stalling
that was not present with \mcasold.
With an aggressive pitch-up control from the pilot, the MCAS
system will not start recovery until {\em after} it detects
a stall is occurring, i.e., the AoA exceeds $\sim17^\circ$.
As a result, the flight will still lose some altitude such
as in \autoref{fig:prestudy:pre_crash_mcas_pilot_error}
while \mcasold{} recovers the aircraft.
On the contrary, \mcasnew{} can only adjust
the HS once, which is not enough for recovery
({\em cf.} \autoref{fig:prestudy:post_crash_mcas_pilot_error}).

However, this evidence
just proves the {\em possibility} and does not explicitly
answer {\em why} or {\em how} the MCAS implementations
directly contribute to the crash of the aircraft. To 
formally investigate these questions, we provide a framework
for defining the error model in \autoref{sec:threat_model},
which incorporates various modes of sensor failure as well
as the timing deadlines that pilots must meet for
aircraft recoverability.




\section{Error Model}
\label{sec:threat_model}


From our previous demonstration, we conclude that
pitch control of the aircraft is fallible through
the input of either the autonomous MCAS or the human
pilot. In this section, we provide a more general
model for the behaviors that can cause these control
failures.
We do not consider errors that are innate to
the flight compute platform itself, such as software
bugs or attack vectors that adversaries may exploit.
We consider these types of errors beyond the scope
of our work, but individual components of the flight
platform may utilize a trusted platform modules
to verify the integrity of the controller
on boot~\cite{microchip:24:tpm}
or mathematical verification~\cite{jeannin:15:emsoft}
to ensure the correctness of the estimators
used in \autoref{alg:sa_mcas}. Moreover, the
estimators used in \autoref{alg:sa_mcas} will
undergo certification by the FAA, so such
safety-critical implementation concerns should
be resolved during this process.



\subsection{Erroneous Sensor Data}
\label{sec:threat_model:sensors}

Like any sensing system, those utilized by an aircraft
may incorrectly measure the environment, and these
sensors can present the erroneous data in a number
of different ways. Here, the models for how these
measurement errors manifest are discussed in
mathematical terms.
We use $\boldsymbol{x}_s(t)$ to denote 
the ground-truth airplane sensor data at time $t$,
and $\overline{\boldsymbol{x}}_s(t)$ to denote
the erroneous airplane sensor data at the same time $t$.
Since we are considering the particular case of MCAS,
we can further simplify $\boldsymbol{x}_s(t)$ to include
the left and right AoA sensor readings.
Our model incorporates Gaussian noise that is 
typically present in sensor measurements as part of $\boldsymbol{x}_s(t)$.

Our selection of sensor errors is representative of
those errors that would occur in the real-world; we
do not consider errors that are theoretically possible
but have no known way of occurring in the real world.
Moreover, we do not consider any combination of the
errors mentioned in this section. In practice,
the more dominant error will have its effects
impact the system, so we consider our evaluation
of the failures in isolation. 

\subsubsection{Sudden Error.}
A {\em sudden error} is defined as
\begin{equation}
    \overline{\boldsymbol{x}}_s(t) = \delta,
\end{equation}
where $\delta$ is a constant value. 
This error is agnostic of the current sensor 
value, making it the most simple sensor error. 
The erroneous data from the AoA 
sensor in ET302 was due to a sudden error,
which may occur due to a jam caused by a bird strike.

\subsubsection{Delta Error.}
For a {\em delta error}, we again assume a
constant value $\delta$ and characterize the error as
\begin{equation}
    \overline{\boldsymbol{x}}_s(t) = \boldsymbol{x}_s(t) + \delta.
\end{equation}
The delta error is simply an offset to the actual sensor
value, in which the constant value $\delta$ is added to 
the existing sensor data. 
The erroneous data from the AoA sensor in JT610 was 
due to a delta error, akin to a miscalibrated sensor.

\subsubsection{Gradual Error.}
The {\em gradual error} is the most sophisticated of
the three, incorporating a function $f(t)$ as part of
error. The gradual error is generalization of the delta
error, replacing $\delta$ with $f(t)$:
\begin{equation}
    \overline{\boldsymbol{x}}_s(t) = \boldsymbol{x}_s(t_0) + f(t).
\end{equation}
Furthermore, in contrast to the delta error,
the $x$-intercept of the function is replaced
with the sensor value at the start of the error, $t_0$.
While the function $f(t)$ can be any function, we
choose a few standard functions: the linear ($f(t) = at$),
quadratic ($f(t) = at^2 + bt$), and logarithmic
($f(t) = a\log(t)$) functions, where $a$ and $b$ are 
predefined coefficients. This error replicates
how a drifting sensor failure may occur over time.


\subsection{Dangerous Pilot Behavior}
The scope of MCAS's authority for counteracting pilot control
is exclusively within the aircraft's pitch axis in the
downward direction. The pilot controls the pitch by
either manually cranking the HS or moving the control 
column to adjust the elevator. See 
\autoref{sec:background:dynamics} for a brief introduction
to longitudinal flight dynamics and control of the aircraft.

For a pilot to provide dangerous control to the aircraft,
we consider two avenues through which this may occur. First,
the pilot could continuously pitch the aircraft up. 
S/he does it by pulling back the control column, which 
in turn commands a consistent input to $\delta_e$ 
in Eq.~(\ref{eqn:shrt_prd}).
Eventually, the aircraft's AoA will exceed $\sim$17$^\circ$, 
causing the aircraft to stall and experience a 
significant decline in altitude before entering a nosedive.

Second, the aircraft may have a major failure that demands
the pilot to respond quickly. During such an event, the FAA
guidance states that a pilot should respond 
within 3s~\cite{gates:20:st}.
However, if the pilot were to take longer to recognize 
and respond to the  occurrence of such an issue, they 
may risk losing control of the aircraft and cause an 
accident. In fact, FAA flight handbooks
refer to reaction times of 4s as 
common~\cite{faa:21:flying_handbook_17}.
A major failure could refer to a correct MCAS
activation that requires the pilot to re-adjust the aircraft
accordingly, or an incorrect MCAS activation that the pilot
must counteract.





\subsubsection{Modeling the Timing Constraint of 
Aircraft Recovery.}

To characterize whether a pilot is performing a timely 
control of an aircraft, we use
\begin{equation}
\tau = \tau_{sensing} + \tau_{action},
\end{equation}
and constrain the success of evading failure with
\begin{equation}
t_{start\ bad\ event} + \tau \leq \tau_{deadline}.
\end{equation}
FAA mandates that a well-trained pilot's sensing of a major
failure, $\tau_{sensing}$, should be 
$<3$ s~\cite{gates:20:st}.
However, we cannot assume the pilot will always be able to
have such quick sensing of a failure, especially if they 
are overwhelmed with other tasks/warnings to attend to.
As previously mentioned, the FAA accepts that it is
common for pilots to require $>4$s to 
react~\cite{faa:21:flying_handbook_17}.

The time it takes a pilot to complete the aircraft 
recovery action from a perceived failure is modeled with
\begin{equation}
\tau_{action} = 
x_{stab\ offset}*\frac{RPD}{RPS}
%
\end{equation}
where we model how long it would take for the pilot 
to move the HS $x_{stab\ offset}$ degrees. 
This movement is dictated by the number of rotations 
per second ($RPS$) the pilot can turn the hand 
crank and the number of
rotations per degree ($RPD$) required to move the HS 
(which is $18$ in a 
737~\cite{stackexchange:19:trim_wheel}).
We note that while the pilot in the physical world may
apply a {\em variable} $RPS$, our investigation
in \autoref{sec:threats} models a
{\em constant} $RPS$ during pilot response to MCAS
events. After recovering from the event, the simulated
pilot will stop rotating the hand crank.

The deadline of aircraft recovery, $\tau_{deadline}$, 
is calculated:
\begin{equation}
\tau_{deadline} = \min \begin{cases}
t_{current} + \frac{h}{v(t+\tau)} \\
t_{last\ MCAS\ fire} + \tau_{MCAS\ c.d.}\end{cases}
\label{eqn:deadline}
\end{equation}
In the first term, we determine whether the altitude, 
$h$, of
the aircraft will reduce to 0 before recovery is completed.
Its catastrophic nature makes it a {\em hard deadline}.
We model the velocity trajectory of the aircraft, $v(t)$,
using the velocity of a falling object with initial
velocity $v_0$:
\begin{equation}
    \frac{dv}{dt} = \frac{1}{m}\sum F(v);\hspace{1em} v_0 < v_t
\end{equation}
mandated by the vertical forces of drag ($D=\frac{1}{2}\rho AC_dv^2$), lift ($L=\frac{1}{2}\rho AC_lv^2$),
gravity ($G=mg$), and thrust ($T$)~\cite{nasa:23:climb}:
\begin{multline}
    \sum F(v) = G + D\sin(c) - L\cos(c) - T\sin(c),
\end{multline}
with climb angle $c$, and the terminal velocity:
\begin{equation}
    v_t = \sqrt{\frac{T\sin(c)-G}{\frac{1}{2}\rho AC_d\sin(c) - \frac{1}{2}\rho AC_l\cos(c)}}.
\end{equation}
obtained by setting $\frac{dv}{dt}=0$, $v=v_t$ and solving for 
$v_t$. In other words, we find the velocity when it is no 
longer changing. Finally, we integrate $\frac{dv}{dt}$ on 
the interval $[v_0,v(t)]$ and find the velocity at time 
$t$~\cite{wiki:24:drag_physics_falling_object}:
\begin{equation}
    v(t) = v_t\tanh\Bigg(\tanh^{-1}\bigg(\frac{v_0}{v_t}\bigg) - \frac{t(T\sin(c)-G)}{v_tm}\Bigg).
\end{equation}

The second term of Eq.~(\ref{eqn:deadline})
relates to the next MCAS activation.
For \mcasold{} and \mcasnew, it offsets the HS
$2.5^\circ$ with a cooldown of 11s if the AoA
remains higher than $17^\circ$;
otherwise, MCAS will not activate again.
During the case of successive MCAS triggers,
the pilot must achieve higher than
$RPD*\frac{x_{stab\ offset}}{\tau_{MCAS\ c.d.}}
= 18*\frac{2.5}{11} = 4.09$ RPS 
of the trim wheel in order to counter the MCAS.
Sustaining a high $RPS$ for a period of multiple
seconds is impractical due to the physical strain
it would cause. Alternatively, during MCAS activation,
the pilot may attempt to physically halt the trim
wheel when MCAS activates, but this similarly
requires the pilot to hold a large amount of
weight for a period of time.

Fortunately, the HS cooldown time is not a
{\em hard deadline}. The pilot failing to undo the HS
displacement once will not lead to catastrophic
failure; it is the result of missing it multiple
times that leads to the catastrophic failure of crashing
the aircraft. This makes the task more similar to
an $(m,k)$-firm guarantee.
When evaluating this aspect of the timing constraints,
we assume $RPD$ and $x_{stab\ offset}$ are static
values built into the design of the aircraft and thus
are not free to change. This assumption is consistent
with the implementation of MCAS in the Boeing 737-MAX.

\section{Semi-Autonomous MCAS (\name)}
\label{sec:design}

Following our preliminary analysis of the differences
between \mcasold{} and \mcasnew{} (\autoref{fig:prestudy}),
we propose Semi-Autonomous MCAS (\name), an MCAS that does
not give static authority to one control
input over the other. Unlike Boeing's MCAS, \name{} uses
a {\em Synthetic Air Data System (SADS) arbiter} to
cross-validate the sensor readings
to first determine whether the pilot or autonomous
control input is correct and then decide which of the two
is allowed to control the pitch of the aircraft.

The SADS is a mechanism that was not originally employed
in the 737-MAX, but it has appeared in advanced commercial
aircraft such as the Boeing 787~\cite{dodt:isasi:11}
and UAVs~\cite{sun:20:thesis}.
It precisely estimates air data that the ADIRU also
supplies, making it an additional source of redundancy.
In the wake of JT610 and ET302 crashes, a U.S.
congressional committee outlined clear evidence that
the use of a SADS in the 737-MAX may have improved its
safety and reliability~\cite{uscongress:20:737maxreport},
but the benefits of its inclusion have neither been shown 
empirically nor has Boeing added one to the 737-MAX.

Moreover, while SADS estimates a sensor's measurement
without directly using that sensor, it may use other sensors' 
measurements that may also have measurement inaccuracies. 
This is a mature tool, so there are many ways
for a SADS to estimate air data~\cite{lie:13:jaircraft}. 
Our novelty is to add an arbiter that is responsible for 
choosing the synthetic data that will be used in the
previously mentioned cross-validation step.
We show that, indeed, SADS presents itself as a promising 
tool in combination with an arbiter.

We note that while \mcasnew{} similarly uses a cross-validation
check to ensure consistency between both AoA sensors, it fails 
to consider instances where both may be incorrect at the same time. 
Functionally, it only checks to see if the measurements of two AoA 
sensors are different from one another by $>5.5^\circ$.
Therefore, if both sensor measurements are incorrect,
yet still similar, the failure is never noticed.
For \name, the incorporation of a SADS arbiter and its
multiple independent estimations of the AoA measurements
ensures that this issue will never arise.

There are a few strategies that can be used after the
cross-validation step determines that the sensor
measurements are incorrect.
This stage of the process is extremely important: we cannot 
employ a strategy that allows the
arbiter to become a new single point of failure. Instead
of directly using the estimated measurement from the SADS,
the arbiter can use the previous data that was determined
correct temporarily. For instance, we can just use the previous
sample directly or extrapolate the correct data using
flight models. However, these strategies may have
unpredictable outcomes without rigorous validation or
if the measurements fail for an extended period of time.
A more predictable strategy
is to drop the
measurements completely. In doing so, the arbiter
no longer becomes a single point of failure; instead,
the arbiter prevents MCAS from making any choice since
it has no data to use.

\begin{algorithm}[t]
\caption{\small SA-MCAS activation using the SADS arbiter technique.}
\label{alg:sa_mcas}
    \footnotesize
    \DontPrintSemicolon
    \SetKwFunction{FArbiter}{arbiter}
    \SetKwFunction{FMCAS}{do\_activate\_SA\_MCAS}
    \SetKwFunction{FADIRUleft}{ADIRU\_left}
    \SetKwFunction{FADIRUright}{ADIRU\_right}
    \SetKwFunction{FSADS}{SADS}
    \SetKwFunction{FIsStall}{is\_stall}
    \SetKwFunction{FWindTriangle}{model\_free\_wind\_triangle}
    \SetKwFunction{FDynamicsModel}{model\_flight\_dynamics}
    \SetKwProg{Fn}{Function}{:}{}
    \Fn{\FMCAS{}}{
        $\boldsymbol{S}_l$, $\boldsymbol{S}_r \gets$
            \FADIRUleft{}, \FADIRUright{}\;
        
        $\boldsymbol{S}_{SADS} \gets$ 
            \FSADS{$S_l$, $S_r$}\;
            
        $\boldsymbol{S}_{correct} \gets$ 
            \FArbiter{$\boldsymbol{S}_l$, $\boldsymbol{S}_r$, $\boldsymbol{S}_{SADS}$}\;
            
        \KwRet \FIsStall{$\boldsymbol{S}_{correct}$}\;
    }
    \;
    \Fn{\FSADS{$\boldsymbol{S}_l$, $\boldsymbol{S}_r$}}{
        $\boldsymbol{w} \gets$ 
            \FWindTriangle{$\boldsymbol{S}_l$, $\boldsymbol{S}_r$}\;
            
        $\boldsymbol{m} \gets$
            \FDynamicsModel{$\boldsymbol{S}_l$, $\boldsymbol{S}_r$}\;
        $\boldsymbol{S}_{SADS} \gets \emptyset$\;
        \Comment{Internal SADS consistency check.}
        \For{$s_w\in\boldsymbol{w}$, $s_m\in\boldsymbol{m}$
            \text{\rm same sensor}}{
            \If{$|s_w - s_m| < \varepsilon$}{
                $\boldsymbol{S}_{SADS} \gets {S}_{SADS} \cup s_w$\;
            }
        }
        
        \KwRet $\boldsymbol{S}_{SADS}$\;
    }
    \;
    \Fn{\FArbiter{$\boldsymbol{S}_l$, $\boldsymbol{S}_r$, $\boldsymbol{S}_{SADS}$}}{
        $\boldsymbol{S}_{correct} \gets \emptyset$\;
        \Comment{External SADS consistency check.}
        \For{$s_l\in\boldsymbol{S}_l$, $s_r\in\boldsymbol{S}_r$, $s_{SADS}\in\boldsymbol{S}_{SADS}$ \text{\rm same sensor}}{
            \If{$|s_l - s_{SADS}| < \varepsilon$}{
                $\boldsymbol{S}_{correct} \gets {S}_{correct} \cup s_l$\;
            }\ElseIf{$|s_r - s_{SADS}| < \varepsilon$}{
                $\boldsymbol{S}_{correct} \gets {S}_{correct} \cup s_r$\;
            }
        }
        \KwRet $\boldsymbol{S}_{correct}$\;
    }
\end{algorithm}

{\bfseries Algorithm and Deployment.}
The full algorithm for SA-MCAS activation is presented
in \autoref{alg:sa_mcas}. To ensure SA-MCAS does
not become a new single-point-of-failure, it
is built with two layers of consistency checking.

First, there is an internal consistency check in {\tt SADS()}, 
which follows the estimation of air data using the
model-free~\cite{klein:06:aiaa} and flight dynamic
model-based~\cite{lie:13:jaircraft,zeis:88:thesis}
methods,
{\tt model\_free\_wind\_triangles} and
{\tt model\_flight\_dynamics}, respectively.
Because there
is a diversity in estimation methods, there
are duplicate estimations of the same air data
measurements from these two methods.
For example, the AoA ($\alpha$)
can be estimated with the model-free method
$\alpha = \tan^{-1}(\frac{u}{v})$ and with the
model-based method $\alpha = f(C_L, M, h)$.
The goal of the internal consistency check
is to ensure that potentially erroneous ADIRU
sensors involved in the estimation methods
are not impacting the final estimation.
The input to {\tt SADS()} includes the left
ADIRU sensor data, $\boldsymbol{S}_l$, and the
right ADIRU sensor data, $\boldsymbol{S}_r$. 
Its output is the set of estimated sensor data,
$\boldsymbol{S}_{SADS}$.

On the other hand, determining which air
data measurements are incorrect is left to
the second, external consistency
check, which occurs during the {\tt arbiter()} step.
If the {\tt SADS()} estimate is
$\varepsilon$ distance away from
the physical measurement of the left or right ADIRU,
the measurement is passed on to the final step.
If the difference exceeds $\varepsilon$, the
measurement is dropped. In this work, the
value of $\varepsilon$ is selected based on
the typical Gaussian  noise of each sensor
measurement. A tight bound is selected to
err in favor of false-negatives since
the effects are in the spirit of the design choices
of Boeing aircraft.
For more conservative
bounds, we would recommend utilizing conformal
prediction for creating a dynamic uncertainty
quantification of the estimated sensor
states~\cite{yang:23:cdc}.
The input to {\tt arbiter()} includes the left
and right ADIRU sensor data, $\boldsymbol{S}_l$
and $\boldsymbol{S}_r$, respectively, and
the {\tt SADS()}
estimated sensor data, $\boldsymbol{S}_{SADS}$.
The output is the set of sensor data that passes the 
external consistency check, $\boldsymbol{S}_{correct}$,
meaning the sensor data that is similar enough to the
estimates.

In the final step of the algorithm, the air data
passed on, $\boldsymbol{S}_{correct}$,
is used to check whether a stall is occurring
or whether the airplane is at risk of stalling. 
If it is, \name{} activates control on the HS.
Similar to Boeing's implementation of MCAS, \name{} is
deployed on the flight control computer of the
737-MAX. This was previously shown in
\autoref{fig:system}; as seen in the figure,
no modification to the communication architecture
is necessary to incorporate \name.




{\bfseries Challenges.}
This study is the first public exploration of
the consequences of the designs of the
MCAS's ability to recover under faults.
As a result,
during the development of \name{} we encountered several
challenges that lead to the primary contributions of this paper. 
We raise the following technical questions:
\begin{itemize}
  \item[\ding{182}] How can we streamline the design and
    evaluation of MCAS
    without a physical aircraft?
    (\autoref{sec:sim})
  \item[\ding{183}] Which control inputs from MCAS and
    the human pilot threaten the safety of the aircraft?
    (\autoref{sec:threats})
  \item[\ding{184}]
    Does \name{} mitigate the issues present
    in \mcasold{} and \mcasnew, and does it satisfy the
    timing constraints for recovering the aircraft?
    (\autoref{sec:prevention})
\end{itemize}

\section{MCAS Simulation}
\label{sec:sim}
This section addresses \textsc{Challenge}$-$\ding{182}.
Using a real airplane as a testbed for evaluation of \name\ is 
unrealistic/infeasible due to the high cost of purchasing the 
aircraft, renting or building a storage facility, 
and hiring pilots. Moreover, our analysis demands us to stress
the limits of the aircraft and put it into hazardous situations
that may ultimately crash it. Thus, the natural solution is
to employ a widely-used flight simulation engine. 
Aerospace companies have custom flight simulators for testing 
their internal products, but they are usually unavailable to 
researchers. As a result, open-source flight simulators, 
such as JSBSim~\cite{berndt:04:aiaa}, are popular among
academic researchers. In particular, JSBSim has
been vetted by NASA, validating its accuracy of
modeling real flight maneuvers~\cite{jackson:15:aiaa}. 
The JSBSim flight simulator enables us to accurately 
model the Boeing 737-MAX's flight and control dynamics.
Furthermore, because of the ease of modeling control loops in 
MATLAB Simulink, an integration of JSBSim to MATLAB was developed 
for this purpose~\cite{sikstom:21:thesis}. 
However, its functionality was limited to just a few hard-coded 
control inputs and no account for pilot control or autonomous
systems.

To overcome this inflexibility, we made an 
extension to the JSBSim Simulink module, which includes several 
user-definable features. Our extension enables the user 
to select any flight sensor input/output to/from JSBSim, 
provides a pilot simulation module with customizable scripts 
for controlling the aircraft, and an MCAS module for 
easy integration of new MCAS designs.
Furthermore, switching between scripts and different MCAS
designs is configurable before simulation execution,
allowing for automated simulation runs without 
any additional manual effort.
Where possible, we have merged features into JSBSim,
while other features specific to our toolkit
are provided as a separate GitHub repository
(see \autoref{fn:github}).

In addition to these features, we provide a module
for injecting sensor errors into the JSBSim sensor
data. This module is capable of injecting the three different
types of erroneous data in \autoref{sec:threat_model:sensors}.

\subsection{Simulation Creation Process}
\label{sec:sim:criteria}
\begin{table}
    \centering
    {\footnotesize
    \begin{tabular}{|| c | c | c ||} 
        \hline
        \rowcolor{black!10}
        \small\bf Maneuver & \small\bf Performed in Crash & \small\bf Ref. \\
        \hline\hline
        Accelerate & $\bullet$ & N/A \\
        \hline
        Climb & $\bullet$ & \cite{krepelka:22:b737800_takeoff} \\
        \hline
        Descend &  & \cite{krepelka:22:b737800_takeoff} \\
        \hline
        Level-Turn &  & \cite{pprn:10} \\
        \hline
        Climb-Turn &  & \cite{krepelka:22:b737800_takeoff,pprn:10} \\
        \hline
        Descend-Turn &  & \cite{krepelka:22:b737800_takeoff,pprn:10} \\
        \hline
        Holding Pattern &  & \cite{faa:22:holding_rules} \\
        \hline
        Takeoff & $\bullet$ & \cite{williams:21:b737800_takeoff,krepelka:22:b737800_takeoff} \\ 
        \hline
        Landing &  & \cite{krepelka:22:b737800_takeoff} \\
        \hline
    \end{tabular}
    }
    \caption{\small Simulated flight maneuvers. We denote those that occur during the crashes of JT610 and ET302.}
    \label{tbl:sim}
    \vspace{-0.1in}
\end{table}

Next we describe how to create simulations in
our toolkit.

{\bf Step 1: Initializing the input/output parameters.}
Before designing the rest of the simulation,
the user must specify the air data they want provided
throughout the simulation. This air data may be flexibly
used and changed in other simulation components.
To request the air data, the user provides an XML file
with the simulator directory paths in which JSBSim stores each
air data measurement. The user also initializes
the erroneous sensor measurements with the time intervals
they occur and the characteristics of the errors as
defined in \autoref{sec:threat_model:sensors}. This
is to be defined within a JSON file.
    
{\bf Step 2: Building the MCAS module.}
We next integrate a specified MCAS design into
the simulation. Using output parameters from the
previous time step (which also may have been
altered by the measurement error module),
we define the specific conditions for
MCAS activation behavior.
For our implementations of Boeing's versions of MCAS, 
we refer to publicly available materials to develop 
a best-effort replica. For instance, 
we refer to~\cite{gates:20:st} for \mcasold{} 
and~\cite{boeing:19:mcas_updates} for \mcasnew. 
We additionally refer to sources such as~\cite{faa:20:faa_mcas_fix,lemme:21:mid_value}
to replicate finer details. Our implementation
is without access or knowledge of proprietary
information that has not been made available publicly.
    
{\bf Step 3: Scripting the pilot behavior.}
Finally, we provide several pilot flight maneuvers as part
of the toolkit, such as takeoff, landing, and turning.
To create a flight maneuver module, it takes the initialized
sensor measurements as input and monitors
them for user-defined conditions that trigger
the next step for the maneuver.
Since this is a simulated pilot, it lacks some of the finer 
feel and touch of a real pilot. 
However, we design these maneuvers using aircraft manuals 
that make suggestions for typical choices in flight 
(see the references in \autoref{tbl:sim}).
While environmental conditions impact some aspects of
each flight maneuver (such as the timing of a specific step),
the steps pilots follow are usually the same.
We validate our simulation of these flight maneuvers in the
following subsection.

\subsection{Example Simulation Scenarios}
\label{sec:sim:result}
\begin{figure*}
    \centering
    \begin{subfigure}[b]{0.25\textwidth}
        \centering
        \includegraphics[width=\textwidth]{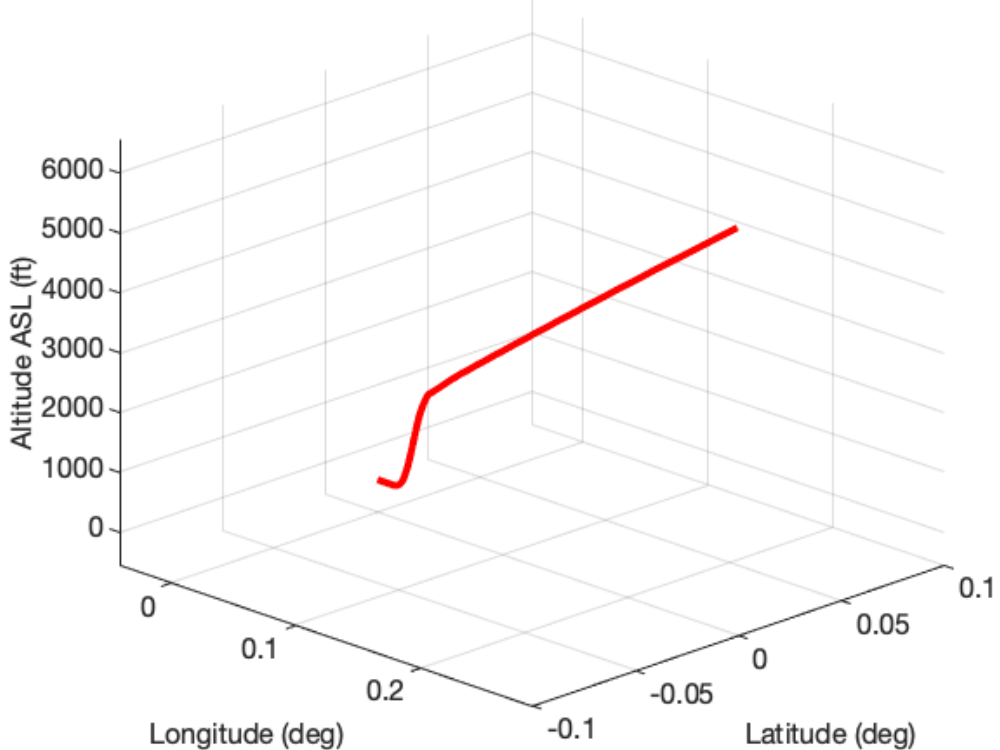}
        \caption{\footnotesize Takeoff.}
        \label{fig:maneuver:takeoff}
        \vspace{-0.08in}
    \end{subfigure}
    \hfill
    \begin{subfigure}[b]{0.25\textwidth}
        \centering
        \includegraphics[width=\textwidth]{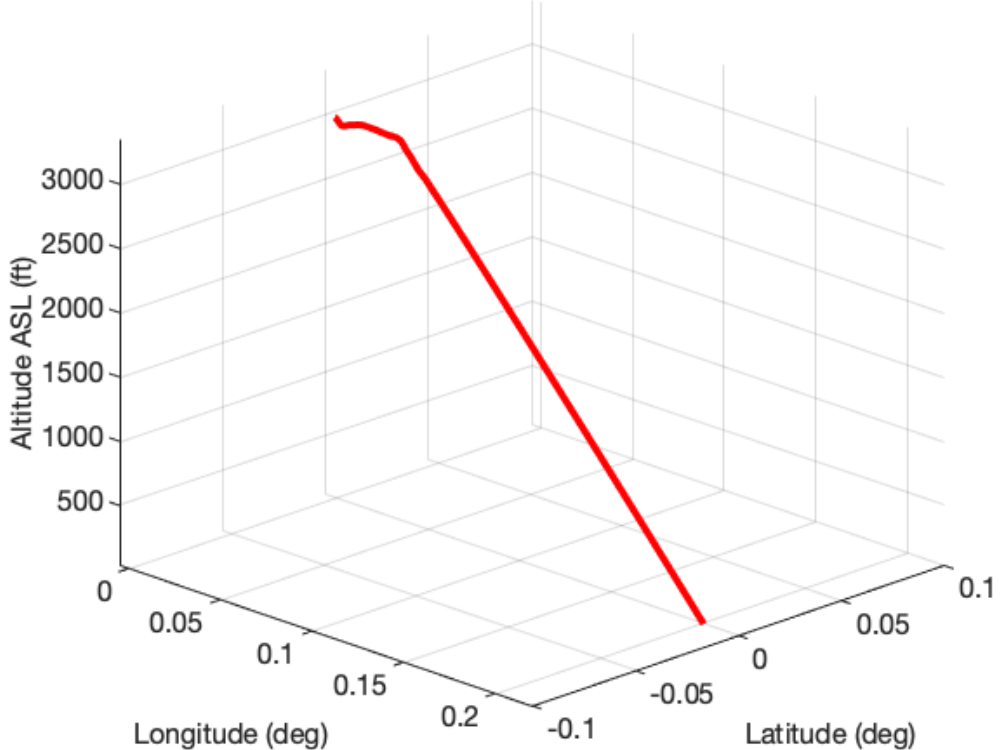}
        \caption{\footnotesize Landing.}
        \label{fig:maneuver:landing}
        \vspace{-0.08in}
    \end{subfigure}
    \hfill
    \begin{subfigure}[b]{0.25\textwidth}
        \centering
        \includegraphics[width=\textwidth]{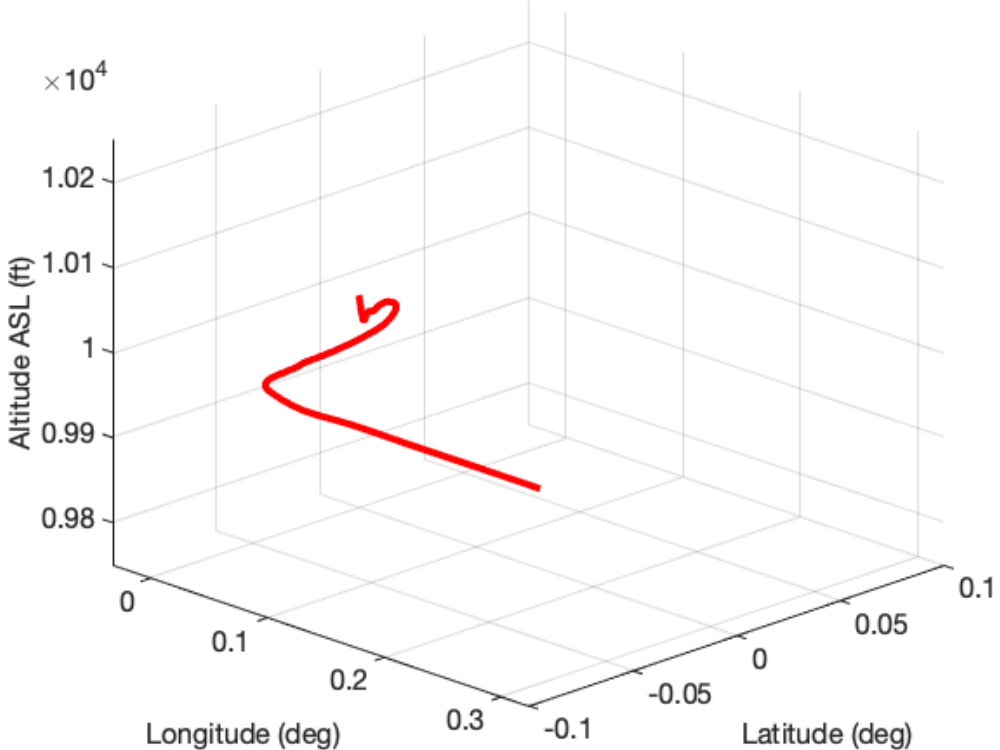}
        \caption{\footnotesize Level turn.}
        \label{fig:maneuver:turn}
        \vspace{-0.08in}
    \end{subfigure}
    \caption{\small Simulated aircraft traces of typical Boeing 737-MAX maneuvers.}
    \label{fig:maneuver}
    \vspace{-0.3in}
\end{figure*}
\begin{table}
    \centering
    {\footnotesize
    \begin{tabular}{|| c | c | c ||} 
        \hline
        \rowcolor{black!10}
        \bf \small & \bf \small Parameter & \bf \small Range \\
        \hline\hline
        \parbox[t]{2mm}{\multirow{4}{*}{\rotatebox[origin=c]{90}{Takeoff}}}
 & Liftoff Indicated Airspeed (kts) & [160, 200]\\
         & Transition to Cruise-Climb Altitude (ft) & [2000, 4000] \\
         & Cruise-Climb Indicated Airspeed (kts) & [220, 300] \\
         & Level-Off Altitude (ft) & [5000, 15000] \\
        \hline
        \hline
        \parbox[t]{2mm}{\multirow{3}{*}{\rotatebox[origin=c]{90}{Landing}}} & Initial Altitude (ft) & [1000, 5500] \\
         & Descent Rate (ft/minute) & [600, 960] \\
         & Final Approach Indicated Airspeed (kts) & [130, 150] \\
        \hline
    \end{tabular}
    }
    \caption{\small Parameter ranges used for validation of maneuvers.}
    \label{tbl:sim_ex}
    \vspace{-0.3in}
\end{table}

Internal states from the JSBSim flight simulator
are fed into our pilot-emulation toolkit, which
makes decisions and generates control commands
based on the thus-provided states.
This setup simulates the pilot 
controlling the plane towards a high-level
goal while considering the flight conditions.

Several flight scenarios (\autoref{tbl:sim})
are used to verify
that the overall simulation functions as expected.
Takeoff and landing scenarios are targeted for
closer study, as these were the scenarios in
which real-world MCAS-related mishaps occurred.
We demonstrate the trace of the flight path for
a few of these scenarios in \autoref{fig:maneuver}.

225 takeoff simulations
and 100 landing simulations
were conducted. Flight parameters were varied
in each simulation (summarized in 
\autoref{tbl:sim_ex}) to cover a broad range
of possible takeoff and landing profiles.
Following this process, we determine that our simulation
of pilot behavior is sufficient for investigating dangerous
flight scenarios.
The flying traces are shown to accurately draw the path of
the desired maneuver. 
Furthermore, we verify the effect of injected
errors to the AoA sensor on the flight path, as seen
in the preliminary study presented in \autoref{fig:prestudy}.

\subsection{Case Study: Simulation of JT610}
\label{sec:threats:case_study}
In order to verify the accuracy of simulating flights
impacted by incorrect sensor data {\em and} to understand
how sensor failures impacted the real flight on JT610, we model
the {\em delta error} that impacted the MCAS
decision-making during JT610. The flight-data
recorder (black box) for JT610 was successfully recovered
by the Indonesian government, and while the raw data was
never publicly released, detailed graphs of the data are
available for analysis~\cite{jt610:18,jt610:18:report}. We use 
the pilot simulation framework described in \autoref{sec:sim}
to model the decisions made by the pilots in JT610,
following the same takeoff procedure. Likewise, we
modeled the same AoA delta error that the left AoA
sensor encountered, which had $\delta \approx 15^\circ$ for the
entire flight from takeoff until crash. Before takeoff,
the error in the left AoA sensor was more variable,
but because it was before takeoff it did not impact the
operation of the airplane. Thus, we do not model this
in our simulation. 

\begin{figure}
    \centering
    \begin{subfigure}[b]{0.45\columnwidth}
        \centering
        \includegraphics[width=\columnwidth]{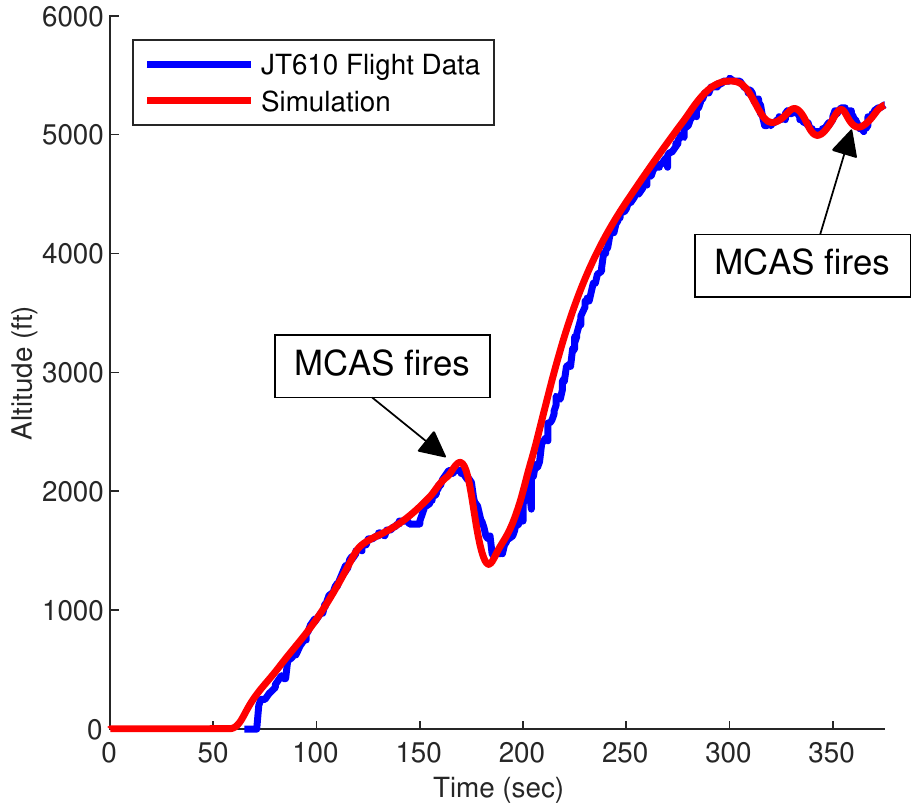}
        \caption{\small Pilot attempts to counter the MCAS misfires,
            similar to the actual flight.}
        \label{fig:casestudy}
    \end{subfigure}
    \hfill
    \begin{subfigure}[b]{0.45\columnwidth}
        \centering
        \includegraphics[width=\columnwidth]{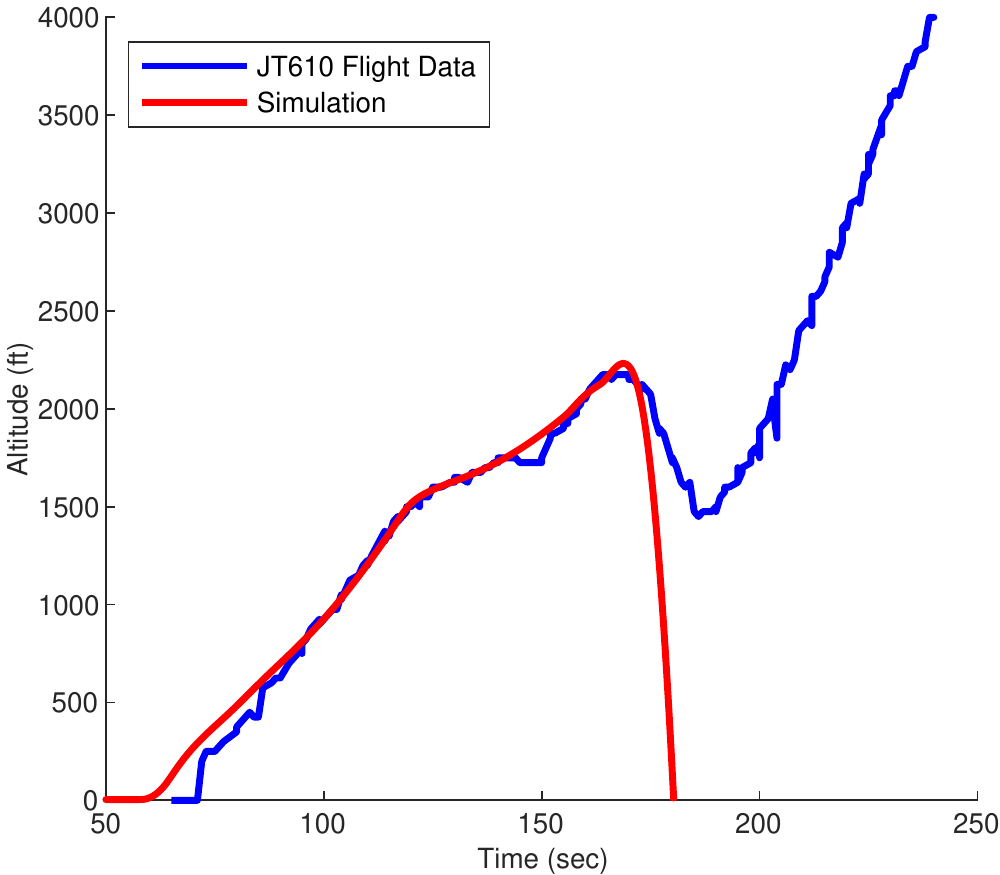}
        \caption{\small Pilot does not intervene at the onset of MCAS
            falsely firing for the first time.}
        \label{fig:casestudy_no_intervention}
    \end{subfigure}
    \caption{\small Simulations of the flight JT610, alongside a delta
        injection with $\delta=15^\circ$. The simulated flights are
        plotted against the flight path of JT610.}
    \label{fig:jt610}
    \vspace{-0.3in}
\end{figure}

As mentioned before, the black box data from JT610
was never publicly released, but we were able to acquire
the flight path of JT610 from Flightradar24~\cite{jt610:18:flightradar}.
We overlay this recovered data with our simulation of
JT610 in~\autoref{fig:jt610}. The overlay on the simulation
demonstrates the capability of our toolkit to accurately
model MCAS misfires in the presence of incorrect sensor values.
In~\autoref{fig:casestudy}, the simulation is shown to closely
overlap with the true flight path of JT610. Since
we are capable of providing an accurate simulation of real
piloting of aircraft experiencing sensor failures, we
provide a deeper investigation of how our proposed error
model from \autoref{sec:threat_model} impacts the simulated
aircraft in the following section.

Before this deeper investigation, our case study
reveals a more interesting pattern
that warrants a closer inspection. The pilot of JT610 was
capable of maintaining an altitude of $\sim$5250 ft.~for
$\sim$7 mins. If an immediate action was not taken by the
pilot, JT610 would have entered a nosedive almost immediately
(simulated in \autoref{fig:casestudy_no_intervention}). In fact,
the pilot recovered the aircraft 21 times in a row before becoming
overwhelmed and handing off responsibility to the co-pilot,
who ultimately failed to recover the aircraft after the hand-off.
Ultimately, the crash of JT610 was a combination of MCAS
{\em repeatedly} activating and the pilot becoming too tired
to manually fight against MCAS automatically trimming the HS.
In other words, this case study underscores the pilot’s capability
of manually recovering an aircraft from {\em rare} false-positive
activation of MCAS.


\begin{tcolorbox}[width=\columnwidth, colback=black!10, arc=3mm]
    \underline{Conclusion for 
    \textsc{Challenge}$-$\ding{182}}:
        \textit{We provide an open-source
        MCAS toolkit built
        on the JSBSim flight simulator and
        MATLAB Simulink. We verify the correctness
        and usefulness of the simulations and include
        guidelines for using this toolkit.}
\end{tcolorbox}

\section{Stress Testing Boeing MCAS}
\label{sec:threats}

\begin{figure*}
    \centering
        \begin{minipage}[b]{1.05\columnwidth}
            \begin{minipage}[b]{0.5\columnwidth}
                \centering
                \subcaptionbox{\small Sudden error flight paths.\label{fig:mcas_old_traj:sudden}}{
                    \includegraphics[width=\textwidth]{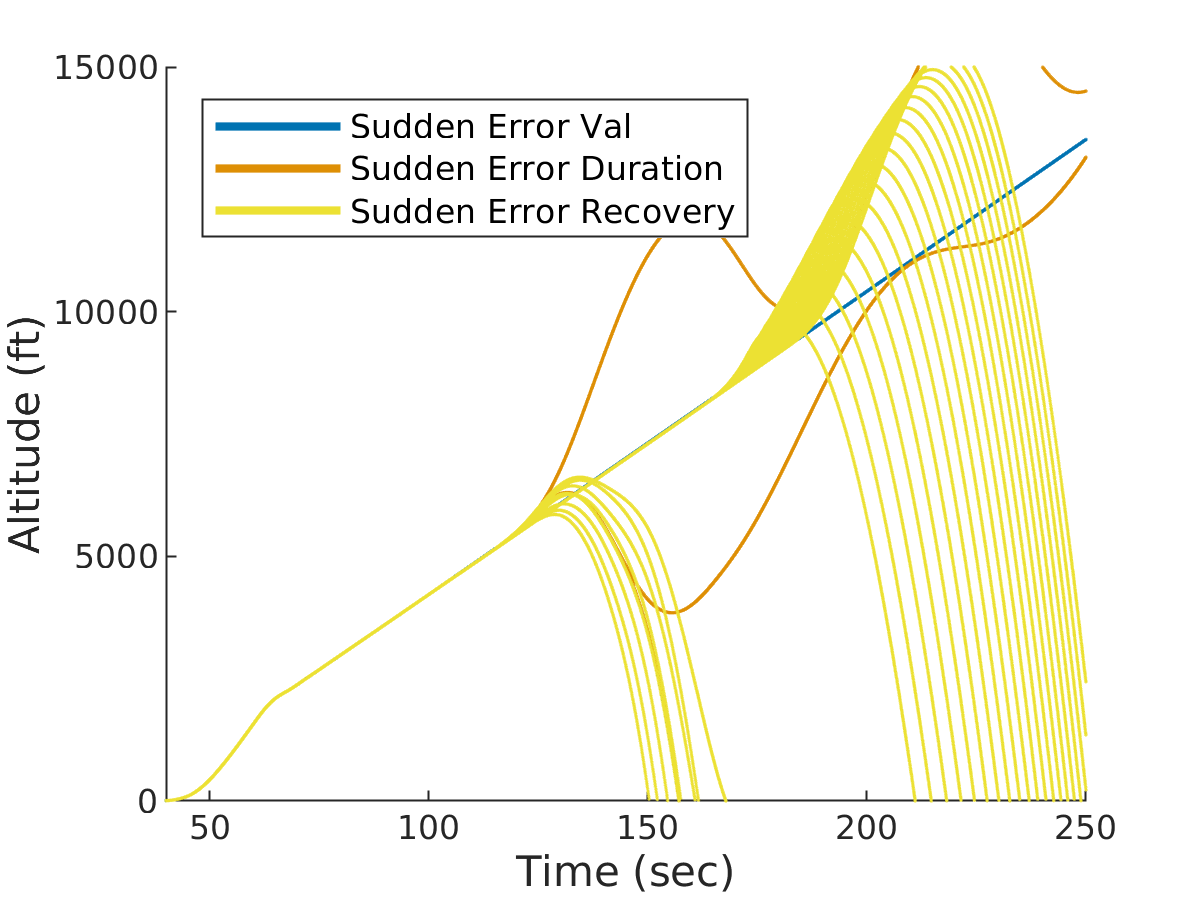}}
            \end{minipage}%
            \hfill
            \begin{minipage}[b]{0.5\columnwidth}
                \centering
                \subcaptionbox{\small Delta error flight paths.\label{fig:mcas_old_traj:delta}}{
                    \includegraphics[width=\textwidth]{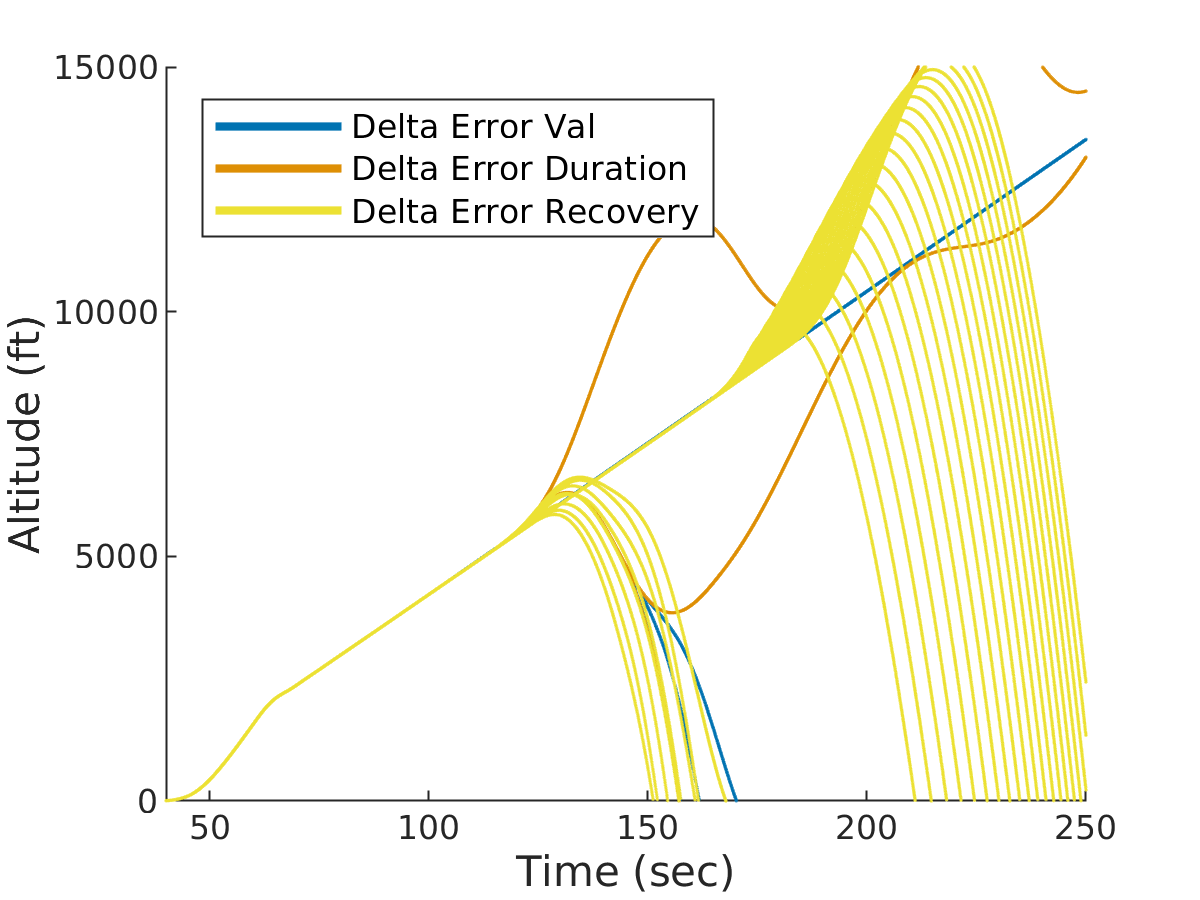}}
            \end{minipage}
            \vfill
            \begin{minipage}[b]{0.5\columnwidth}
                \centering
                \subcaptionbox{\small Gradual error flight paths.\label{fig:mcas_old_traj:gradual}}{
                    \includegraphics[width=\textwidth]{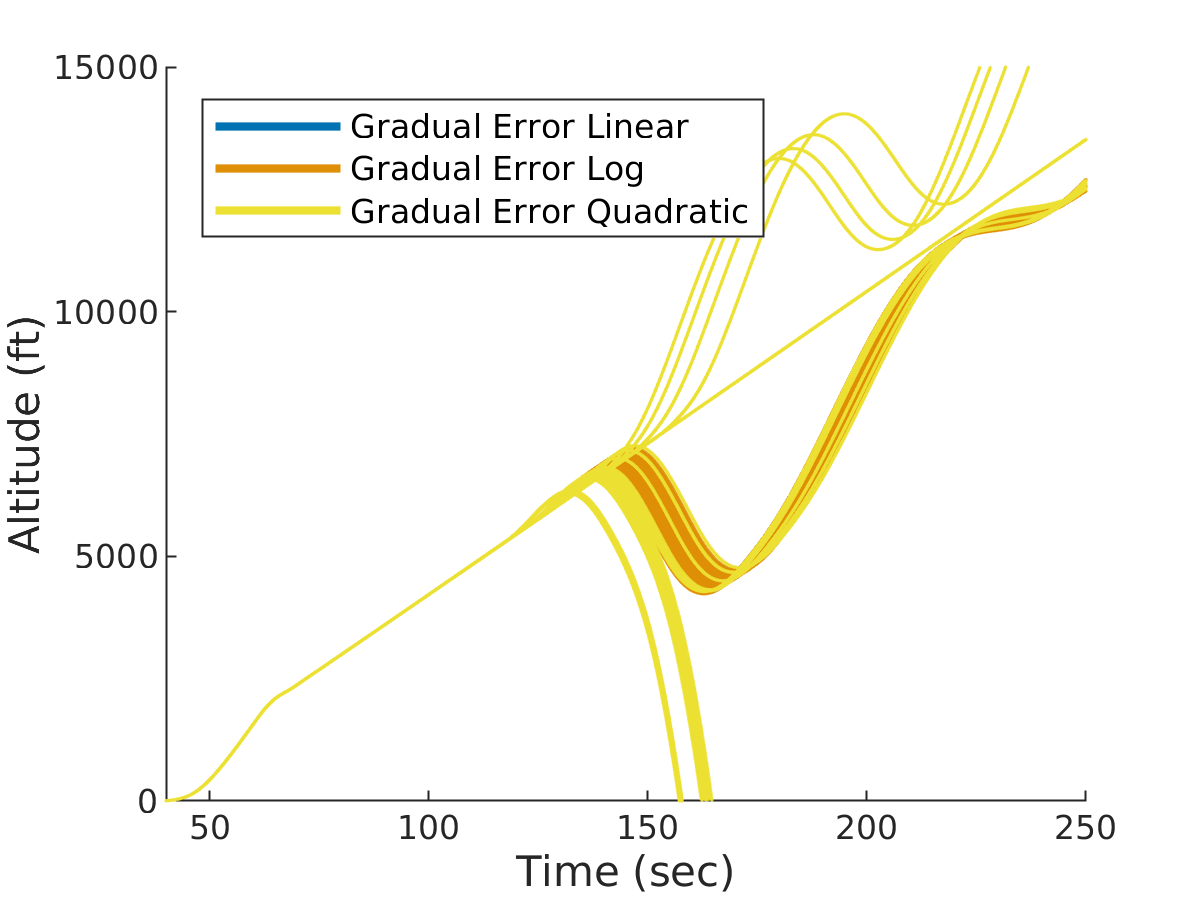}}
            \end{minipage}%
            \hfill
            \begin{minipage}[b]{0.5\columnwidth}
                \centering
                \subcaptionbox{\small Pilot-induced stall flight paths.\label{fig:mcas_old_traj:stall}}{
                    \includegraphics[width=\textwidth]{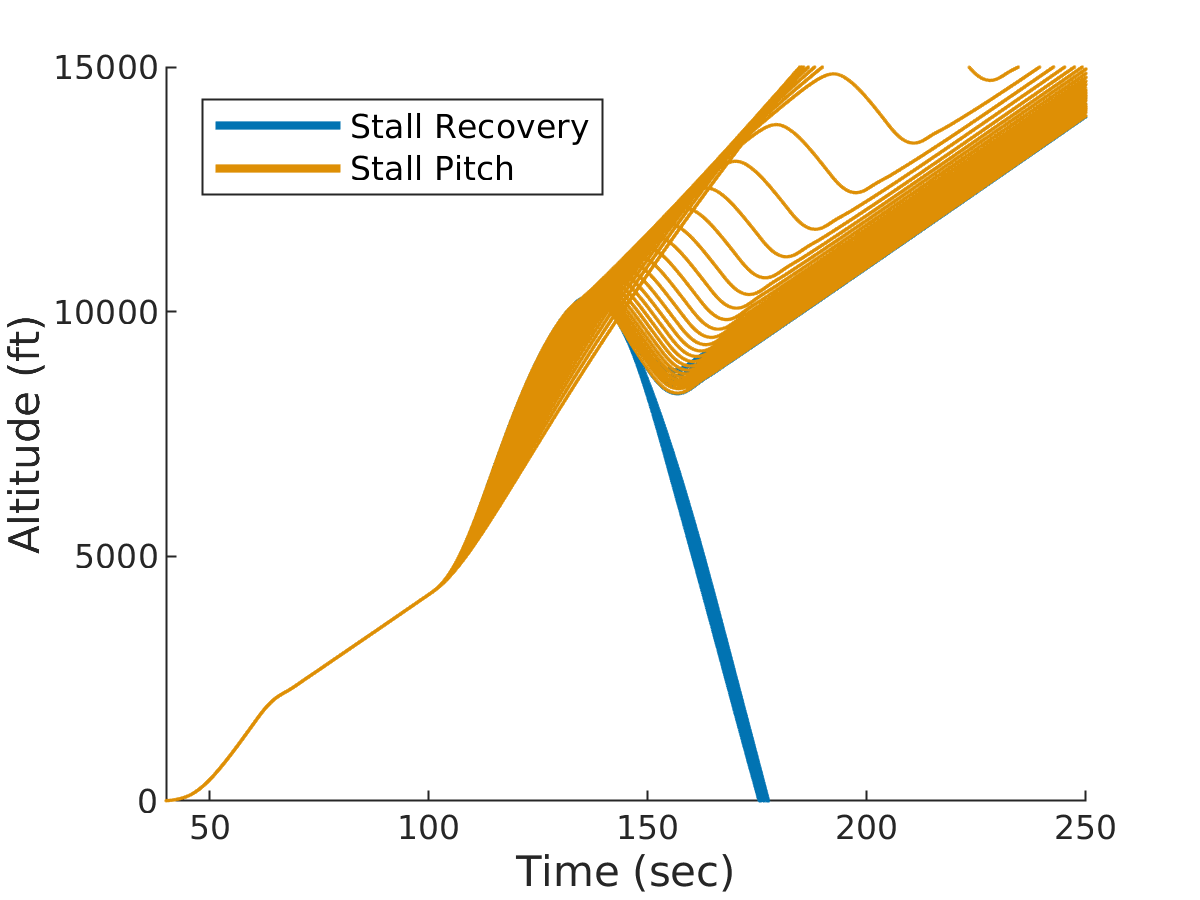}}
            \end{minipage}
        \end{minipage}%
        \hfill
        \begin{minipage}[b]{0.95\columnwidth}
            \begin{subfigure}[b]{\columnwidth}
                \centering
                \includegraphics[width=0.8\textwidth]{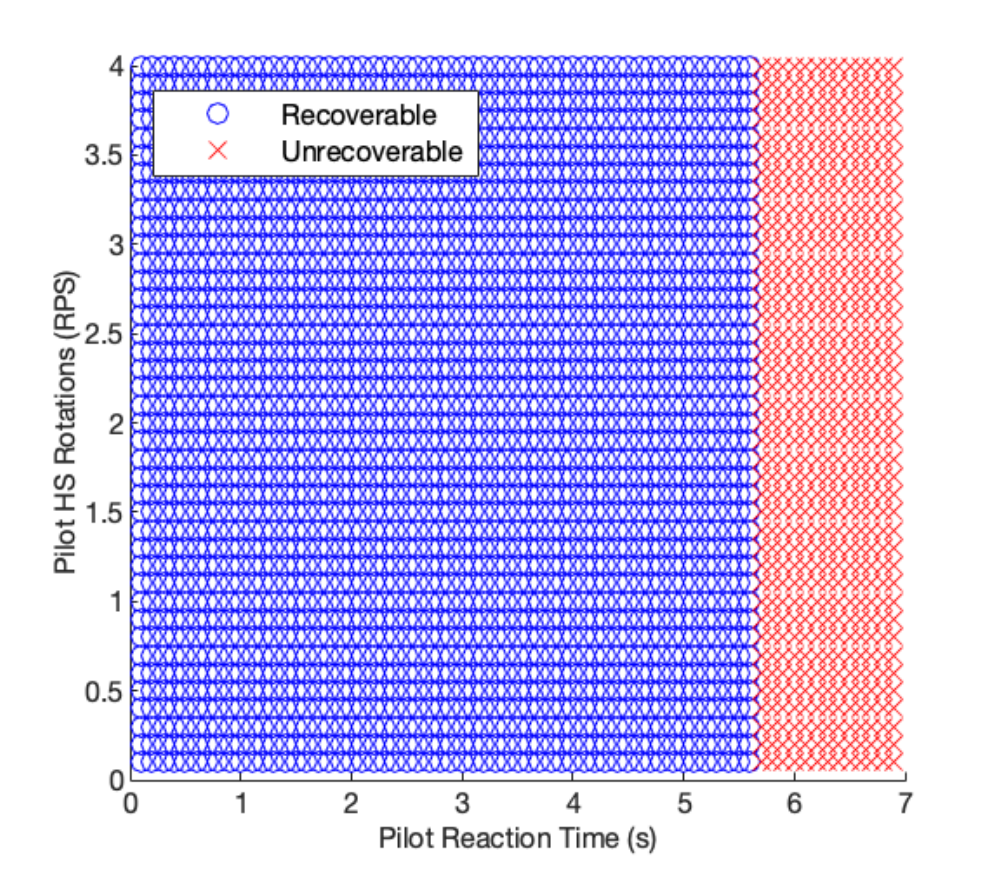}
                \caption{\small Pilot can have
                some variability in their response time and
                exerted effort on the HS hand-crank. This
                figure examines the impact of this variability
                on aircraft recovery. The recoverability of the
                flight {\em is not} dictated by the pilot's reaction
                speed and rotation of the HS.}
            \label{fig:mcas_old_safe_state}
            \end{subfigure}
            \vspace{-0.4in}
        \end{minipage}
    \caption{\small
    Summary of the stress test simulation for \mcasold.}
    \label{fig:mcas_old_traj_safe_state}
    \vspace{-0.2in}
\end{figure*}

Using the erroneous sensor injection tool included 
as part of our simulation toolkit presented in
\autoref{sec:sim}, we try to cause failures on MCAS to
reveal the precise conditions under which dangerous
control of the aircraft is possible.
Moreover, we have the pilot cause a
stall in order to evaluate whether MCAS can mitigate
the dangerous control. Such stress testing incorporates
the error model from \autoref{sec:threat_model}.
Doing so leads to a conclusion for
\textsc{Challenge}$-$\ding{183}.

\subsection{Methodology for Stress Tests}
\label{sec:threats:methodology}
For the {\em sudden} and {\em delta errors}, we conduct 
three stress tests for each error: (1) $\delta\in[0,90]$
for time range $t\in[100,150]$, pilot reacts after 5s;
(2) $\delta = 18^\circ$ for time range $t\in[100,t_{end}]$,
$t_{end}\in[110,180]$, pilot reacts after 5s; and (3)
$\delta=18^\circ$ for time range $t\in[100,150]$, 
pilot reacts $\in[0,10]$s.
For each stress test, we perform a parameter sweep using
binary search in order to find the boundary for which 
the aircraft is no longer recoverable.
The ranges for these test are chosen based on
the physical limitations of the aircraft and pilot.
For instance, while $\delta$ may theoretically be higher
than 90$^\circ$, the aircraft will never pitch high
enough for this to be the case.

For the {\em gradual errors}, we conduct three 
experiments, one for each of linear, quadratic, and
logarithmic functions. The controlled settings for
these stres tests are, respectively, (1) $f(t)=at$ where
$a\in[0,3]$ and pilot
reaction after 5s; (2) $f(t)=a\log(t)$ where
$a\in[0,500]$ and pilot
reaction after 5s; and (3) $f(t)=at^2$ where
$a\in[0,3]$
and pilot reaction after 5s.

The goal of these stress tests is to see whether MCAS
will incorrectly activate
(i.e., actives when no stall is occurring).
When MCAS incorrectly activates, the pilot is providing
a normalized elevator input of -0.1 and their reaction
is to re-trim the HS at a rate of 3.5 RPS.

To incorporate {\em dangerous pilot behavior}, we perform
two separate stress tests. (1) We simulate a pilot pitching
up the aircraft $\in[20,90]^\circ$ in order to cause a stall.
5s after the stall occurs (i.e., MCAS correctly activates),
the pilot begins recovery of the aircraft, which follows a
pitch-down of the aircraft to gain speed, then a pitch-up
followed by trimming the HS. (2) We simulate a pilot's
recovery time in order to test the timing aspect of the
recovery. The pilot pitches up the aircraft 50$^\circ$ to
cause a stall. After MCAS activates, we vary the recovery
reaction time $\in[0, 10]$s.

We provide further investigation into the impact
that the variable pilot behavior may have on the
timing analysis. Our analysis is shown in
\autoref{fig:mcas_old_safe_state}. The $x$ axis
is the pilot reaction time, $\tau_{sensing}$,
in the range of 0.1 to 7s. The $y$ axis is
the component of $\tau_{action}$ that is under
the pilot's control, the $RPS$ of the HS hand-crank,
which is in the 0.1 to 4 RPS range.

\subsection{Stress Test of \mcasold}
\label{sec:threats:mcasold}
\begin{table}
    \centering
    {\footnotesize

    \begin{tabular}{|| c | c | c | c ||} 
        \hline \rowcolor{black!10}
        \textbf{\backslashbox{Stress Test}{MCAS}} & \textbf{\mcasold} & \textbf{\mcasnew} & \textbf{\name}\\
        \hline\hline
        Sudden Val & $17^\circ$ & No failure & No failure \\
        Sudden Duration & $140.5450$s & No failure & No failure \\
        Sudden Recovery & $2.7991$s & No failure & No failure \\
        Delta Val & $13.8750^\circ$ & No failure & No failure \\
        Delta Duration & $140.5450$s & No failure & No failure \\
        Delta Recovery & $2.7991$s & No failure & No failure \\
        Gradual Linear & $1.5000$ & No failure & No failure \\
        Gradual Log & $222.5000$ & No failure & No failure \\
        Gradual Quadratic & $1.4999$ & No failure & No failure \\
        Stall Pitch & $51.5497^\circ$ & $46.2531^\circ$ & $51.5497^\circ$ \\
        Stall Recovery & $5.6333$s & $3.9084$s & $5.6333$s \\
        \hline
    \end{tabular}
    }
    \caption{\small
    Summary table of the results from our stress test of 
    each MCAS. Each row is associated with a particular 
    stress test. Details on these tests are reported in 
    \autoref{sec:threats:methodology}.
    Each entry is the lower bound for failure, and
    the higher numbers are better.
    }
    \label{tbl:mcas_summary}
    \vspace{-0.3in}
\end{table}

The summary of the results for the stress test
of \mcasold{} may be found in
\autoref{fig:mcas_old_traj_safe_state}
and the first column of
\autoref{tbl:mcas_summary}.

\subsubsection{Sudden \& Delta Sensor Errors.}
For the variant of these tests that increase
the measurement error incrementally, the plane was
irrecoverable when the error is large enough
to start triggering \mcasold{}, i.e., the
measurement error causes the AoA value to exceed
17$^\circ$. This is primarily
due to the simulated pilot responding too late.
For the tests that stress the measurement error
duration, the results are similar; after
\mcasold{} activates a third time,
the pilot is unable to recover the aircraft.
This result is consistent with what was
observed in the crashes of JT610 and
ET302 --- after the third activation, \mcasold{}
has displaced the location of the HS
substantially enough to make the aircraft irrecoverable.
Finally, we observe that no matter the response time
of the pilot, the flight cannot be recovered
if a measurement error is sustained
for a long enough period of time.

\subsubsection{Gradual Sensor Errors.}
For the case of the gradual sensor measurement
error, \mcasold{} is only capable of preventing
a crash when the log function and the linear
function are parameterized with $a<222.5$ and $a<1.5$,
respectively. The quadratic case only has successful recovery
when $a<1.5$.
This means that a moderately gradual drift in measurement
error may incorrectly invoke \mcasold{} and cause
the aircraft to be irrecoverable.

\subsubsection{Pilot Stalling.}
Our investigation into pilot stalling reveals that
with the assistance of \mcasold{} the factor most
important is the pilot's reaction time. When
the pilot has a reaction time of 5s, all pitch
angles from 20$^\circ$ to 50$^\circ$ have
recoverable stall events. On the other hand, with
a pitch angle of 50$^\circ$, the stall event is
recoverable when the pilot reacts in at most 5.63s.
After 5.63s, none of the flights are recoverable.

\subsubsection{Deadline \& Timing Analysis.}
The stall recovery test demonstrates that
an \mcasold-assisted recovery of the stall is
only possible when the pilot reacts within
$\sim$5.63s.
We investigate this further in \autoref{fig:mcas_old_safe_state}. 
We find that the pilot reaction time is the main contributor 
to aircraft recovery. Interestingly, it is
inconsequential how much effort the pilot
exerts toward recovery, i.e., the number of
rotations per second on the hand-crank that
adjusts the HS.
Also, notably, the sudden and delta
errors clearly are irrecoverable when the
pilot reacts to the incorrect \mcasold{} activation
too late (e.g., our set 5s for the experiments).
If the pilot responds quickly, recovery is still possible.
The pilot, therefore,
has stricter reaction constraints for safe
recovery of the aircraft when \mcasold{} falsely
activates.


\subsection{Stress Test of \mcasnew}
\label{sec:threats:mcasnew}

\begin{figure*}
    \centering
        \begin{minipage}[b]{1.05\columnwidth}
            \begin{minipage}[b]{0.5\columnwidth}
                \centering
                \subcaptionbox{\small Sudden error flight paths.\label{fig:mcas_new_traj:sudden}}{
                    \includegraphics[width=\textwidth]{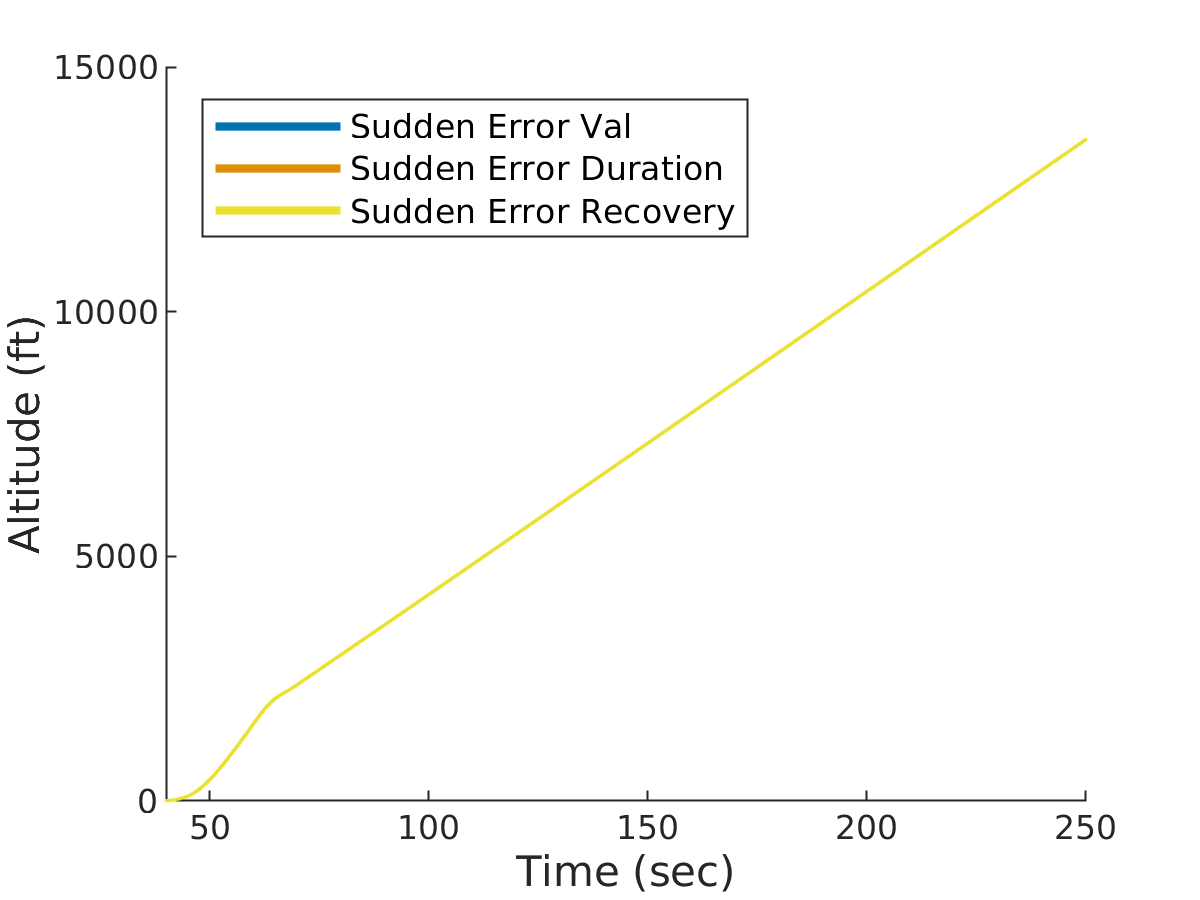}}
            \end{minipage}%
            \hfill
            \begin{minipage}[b]{0.5\columnwidth}
                \centering
                \subcaptionbox{\small Delta error flight paths.\label{fig:mcas_new_traj:delta}}{
                    \includegraphics[width=\textwidth]{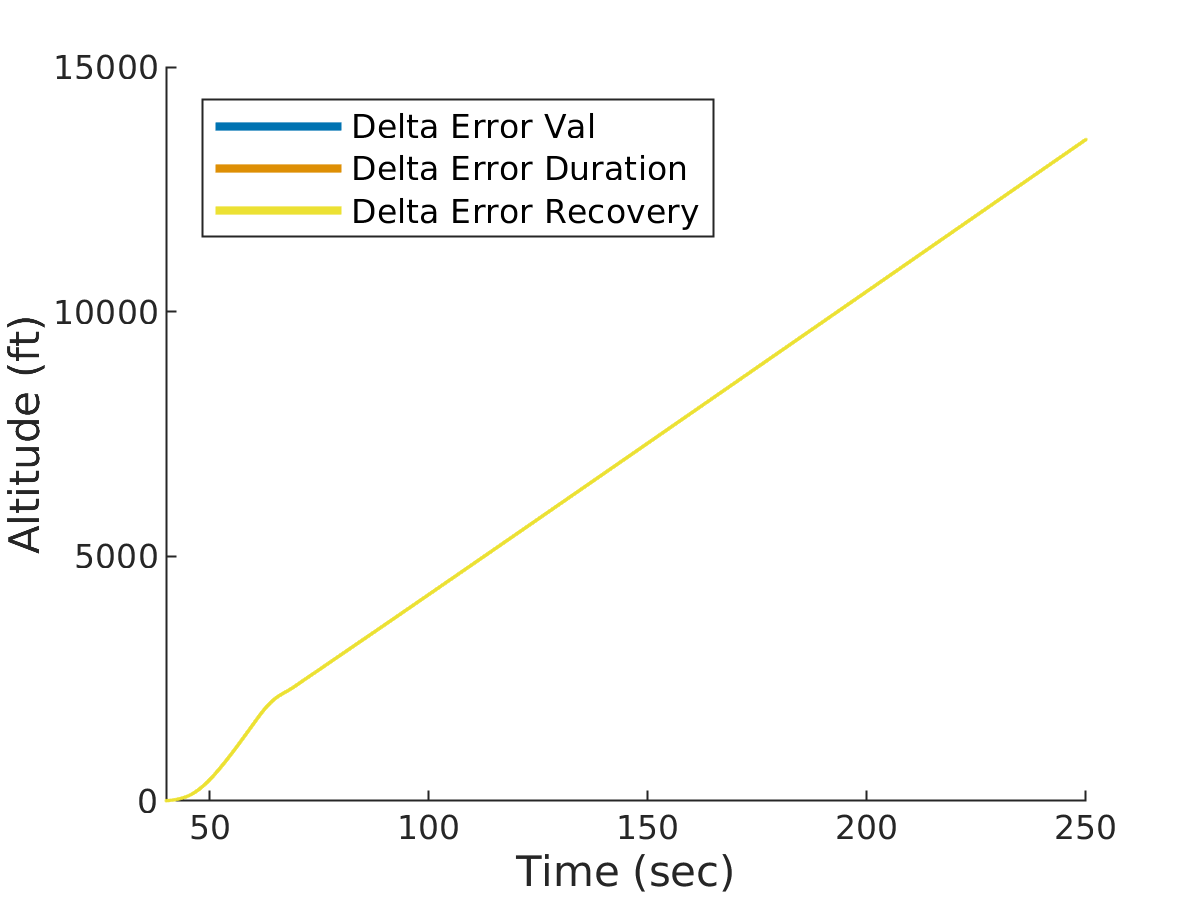}}
            \end{minipage}
            \vfill
            \begin{minipage}[b]{0.5\columnwidth}
                \centering
                \subcaptionbox{\small Gradual error flight paths.\label{fig:mcas_new_traj:gradual}}{
                    \includegraphics[width=\textwidth]{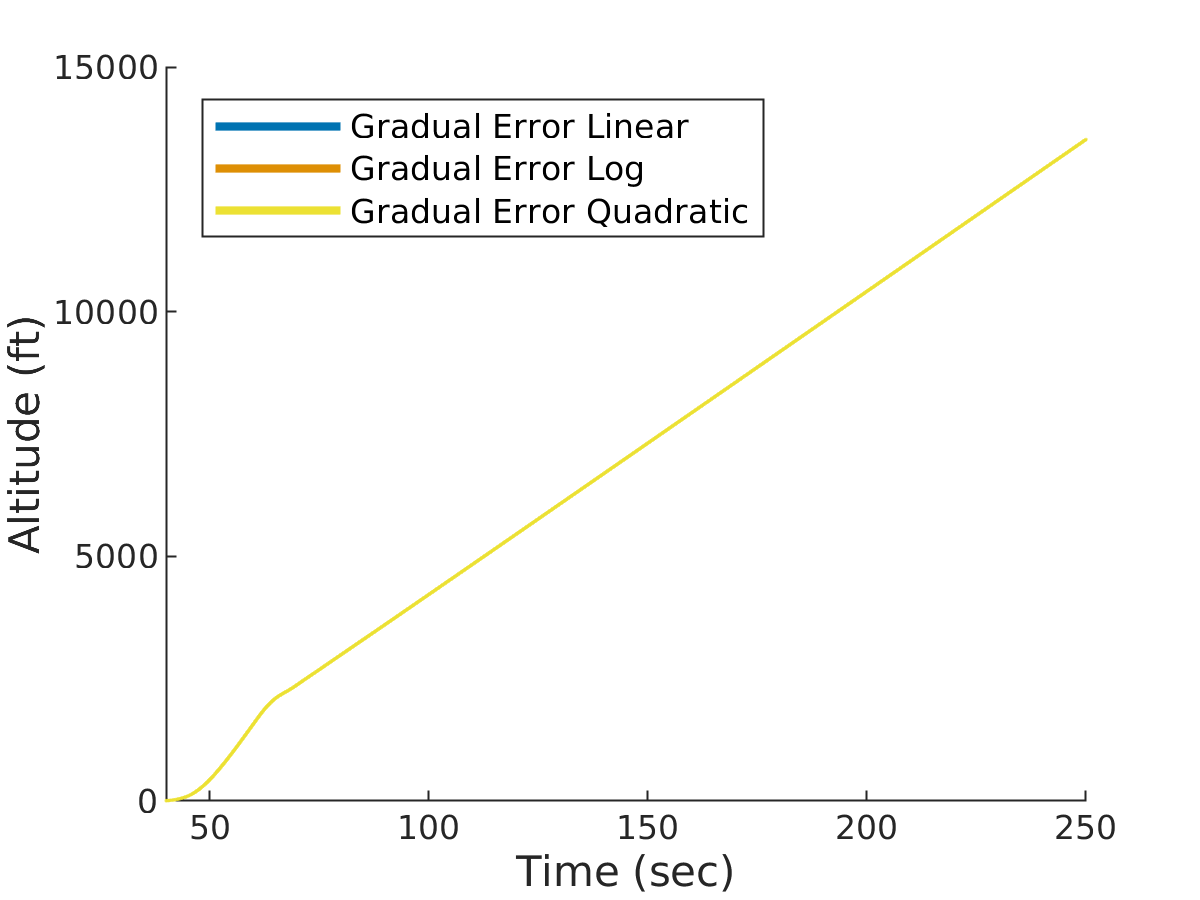}}
            \end{minipage}%
            \hfill
            \begin{minipage}[b]{0.5\columnwidth}
                \centering
                \subcaptionbox{\small Pilot-induced stall flight paths.\label{fig:mcas_new_traj:stall}}{
                    \includegraphics[width=\textwidth]{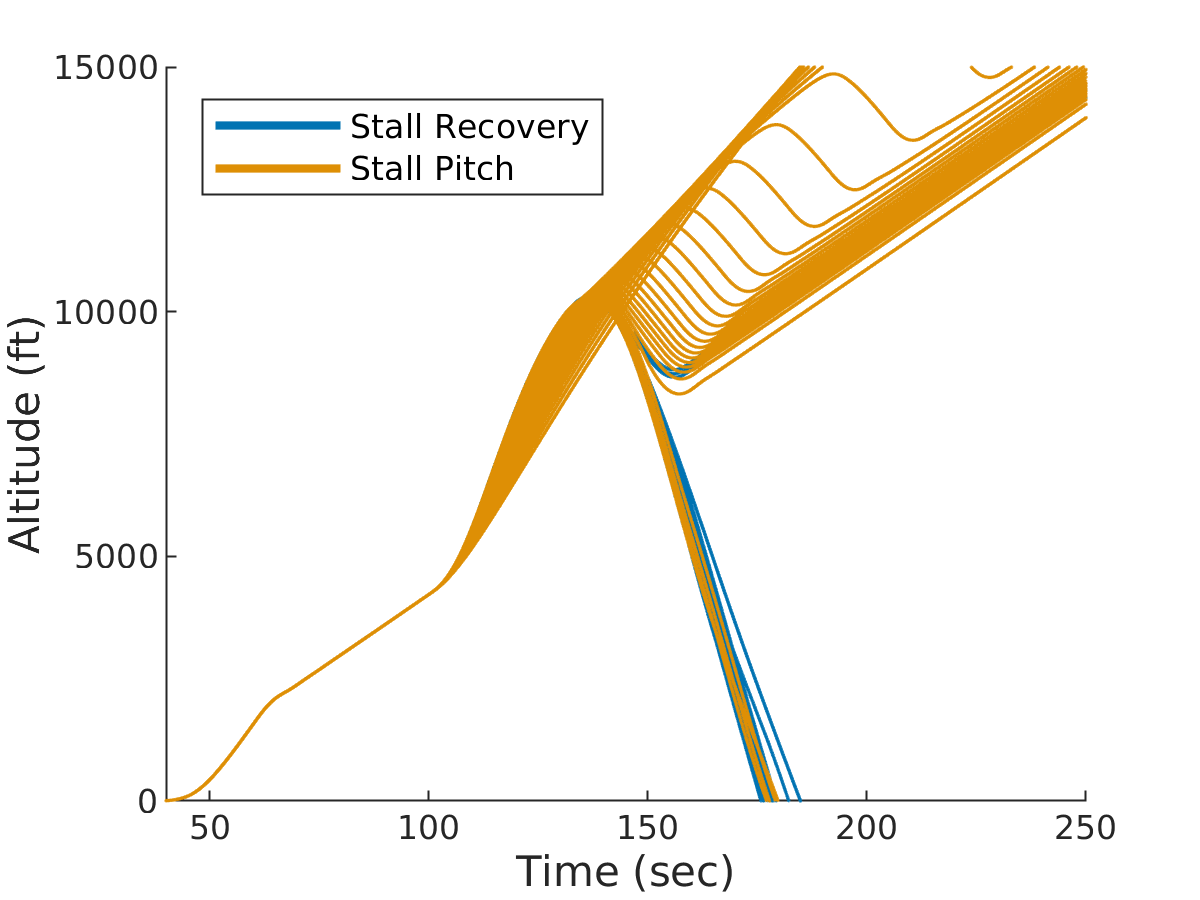}}
            \end{minipage}
        \end{minipage}%
        \hfill
        \begin{minipage}[b]{0.95\columnwidth}
            \begin{subfigure}[b]{\columnwidth}
                \centering
                \includegraphics[width=0.8\textwidth]{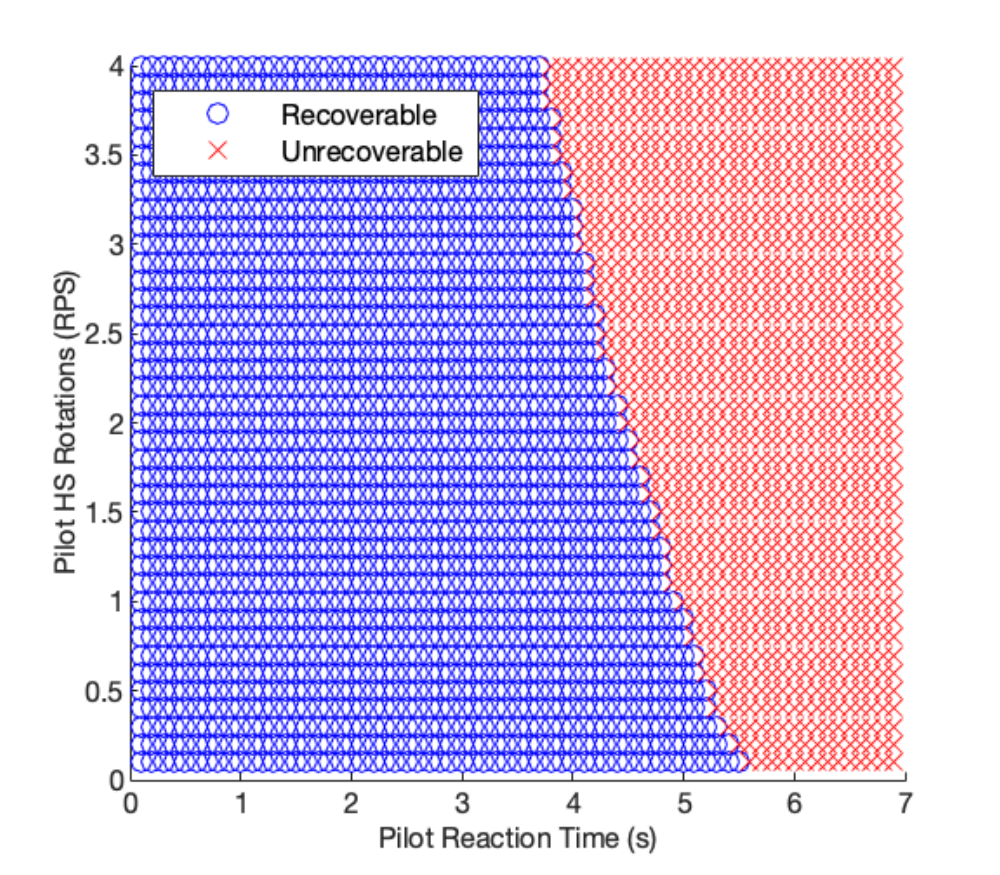}
                \caption{\small Pilot can have
                some variability in their response time and
                exerted effort on the HS hand-crank. This
                figure examines the impact of this variability
                on aircraft recovery. The recoverability of the
                flight {\em is} dictated by the pilot's reaction
                speed and rotation of the HS.}
            \label{fig:mcas_new_safe_state}
            \end{subfigure}
            \vspace{-0.4in}
        \end{minipage}
    \caption{\small 
    Summary of the stress test simulation for \mcasnew.}
    \label{fig:mcas_new_traj_safe_state}
    \vspace{-0.25in}
\end{figure*}

The summary of the results for the stress test
of \mcasnew{} may be found in
\autoref{fig:mcas_new_traj_safe_state}
and the second column of
\autoref{tbl:mcas_summary}.

\subsubsection{Sensor Errors.}
For \mcasnew, all flight paths are recovered
for the sensor measurement error tests.
This is consistent with Boeing's claims and
is unsurprising since our stress tests only cause
measurement error in one of the two AoA sensors.
In other words, just comparing the difference
between the two sensor measurements is sufficient.
However, if both AoA sensors have a similar
measurement error, \mcasnew{} is not sufficient.
Trivially, this is not detectable nor recoverable
by \mcasold{} since it only uses one of the AoA
sensors.
In \autoref{sec:prevention} we show how
\name{} is capable of recovery against this 
class of measurement error.

\subsubsection{Pilot Stalling.}
Consistent with our initial investigation
in \autoref{fig:prestudy}, we find that because
\mcasnew{} is functionally only allowed to
activate once, there are stall events during
a high pitch-up that render the flight
unrecoverable. With \mcasold, these flights
were recoverable because of its repeated
assistance. Moreover, with the fixed pitch
angle set to 50$^\circ$, the stall event
is only recoverable if the pilot responds
in $\sim3.9$s. This is $\sim$31\% decrease in
the total time the pilot had previously to
recover from a similar stall event with \mcasold.

\subsubsection{Deadline \& Timing Analysis.}
For \mcasnew, the amount of time that the pilot
has to respond to a stall event is reduced by
nearly 2s. While there are clear gains from
eliminating the pathway for these errors to
occur, enforcing such restrictions on the
MCAS activation places burdens on the pilot.
We conduct an in-depth analysis of the
pilot's influence toward recovery in
\autoref{fig:mcas_new_safe_state}. The results
are consistent with those we observe in
\autoref{fig:mcas_old_safe_state}. It also
demonstrates that the time the pilot can respond
to recover the aircraft is consistently reduced.



\begin{tcolorbox}[width=\columnwidth, colback=black!10, arc=3mm]
    \underline{Conclusion for 
    \textsc{Challenge}$-$\ding{183}}:
        \textit{We demonstrate a series of control
        threats outside of those that caused the
        original 737-MAX crashes. We also 
        demonstrate that the new Boeing MCAS 
        design is susceptible to the newly 
        identified control threats from the pilot.
        Our analysis unveils precise upper bounds
        for aircraft recoverability during erroneous
        MCAS events.}
\end{tcolorbox}

\section{Evaluation of \name}
\label{sec:prevention}

\begin{figure*}
    \centering
        \begin{minipage}[b]{1.05\columnwidth}
            \begin{minipage}[b]{0.5\columnwidth}
                \centering
                \subcaptionbox{\small Sudden error flight paths.\label{fig:sa_mcas_traj:sudden}}{
                    \includegraphics[width=\textwidth]{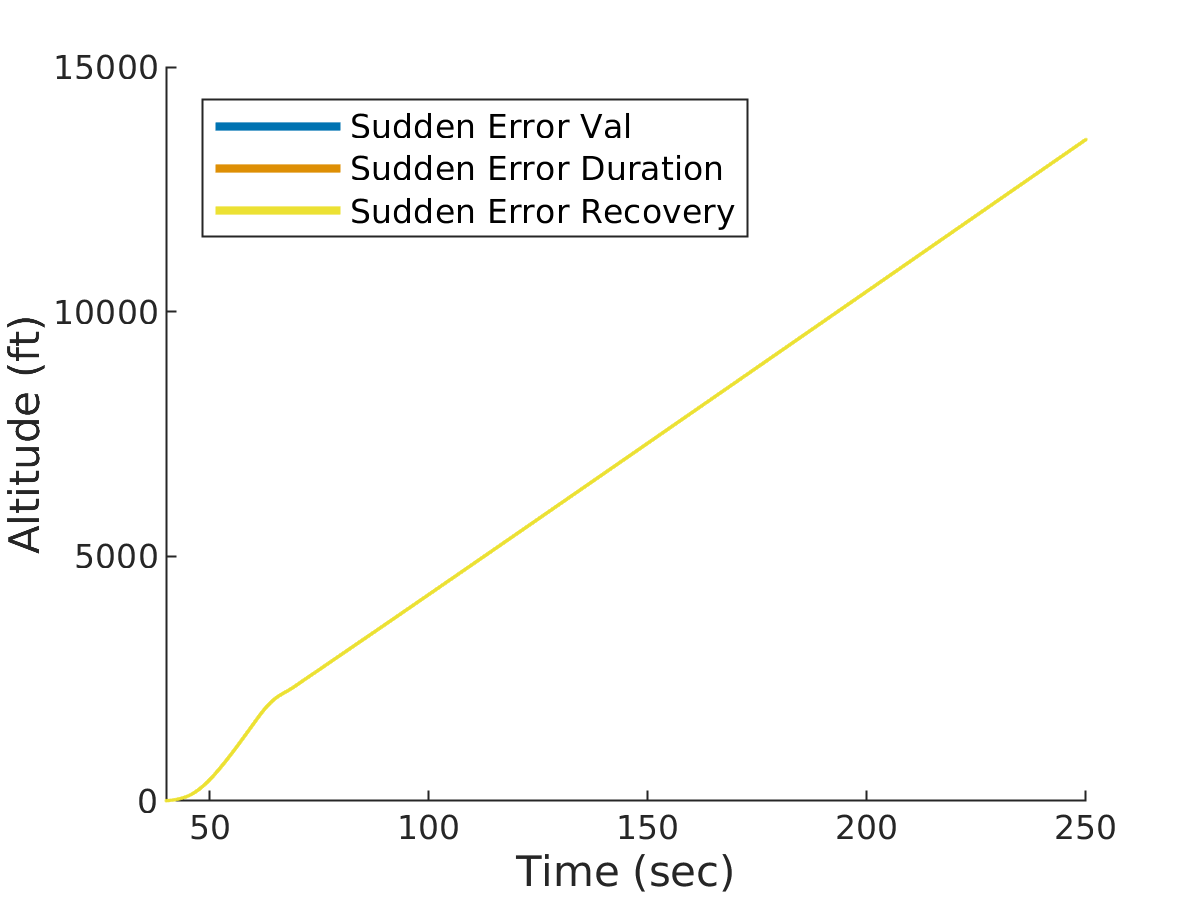}}
            \end{minipage}%
            \hfill
            \begin{minipage}[b]{0.5\columnwidth}
                \centering
                \subcaptionbox{\small Delta error flight paths.\label{fig:sa_mcas_traj:delta}}{
                    \includegraphics[width=\textwidth]{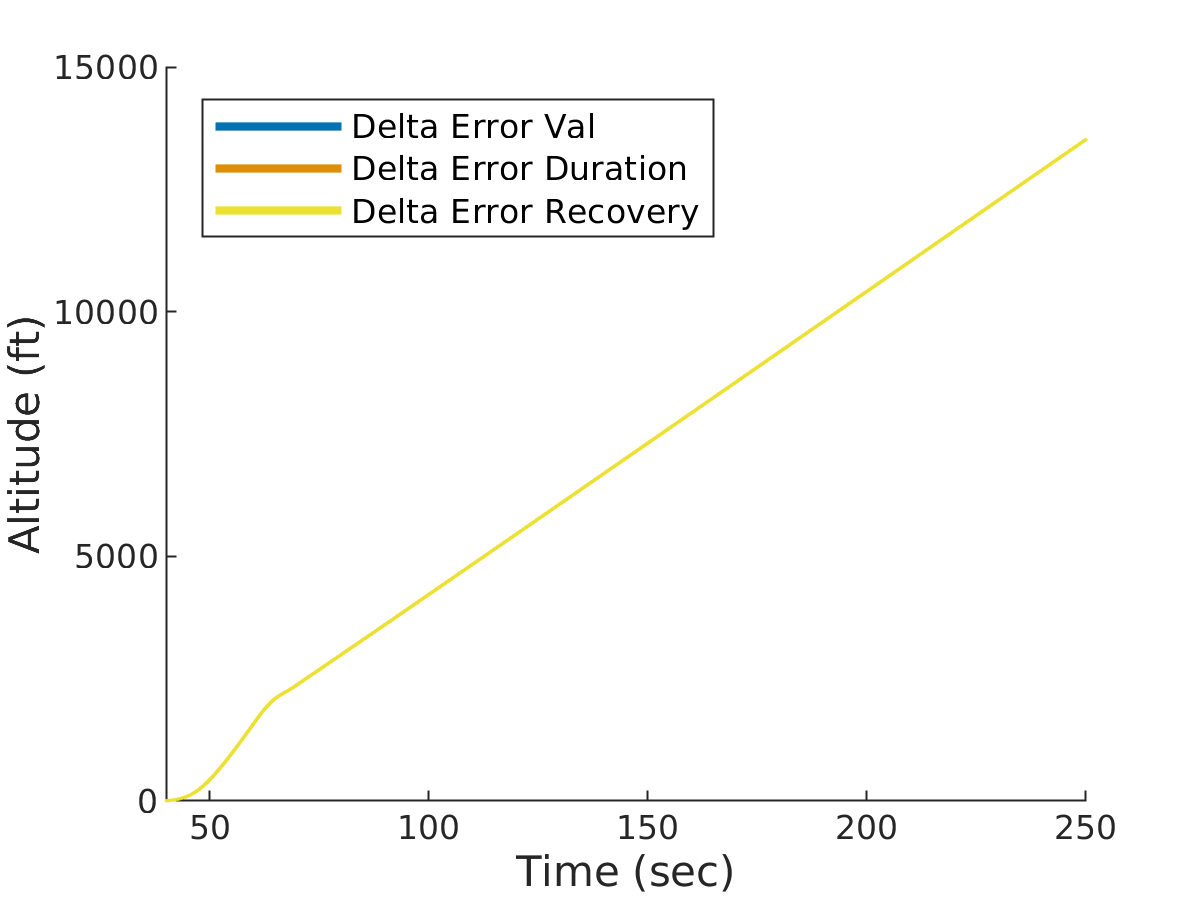}}
            \end{minipage}
            \vfill
            \begin{minipage}[b]{0.5\columnwidth}
                \centering
                \subcaptionbox{\small Gradual error flight paths.\label{fig:sa_mcas_traj:gradual}}{
                    \includegraphics[width=\textwidth]{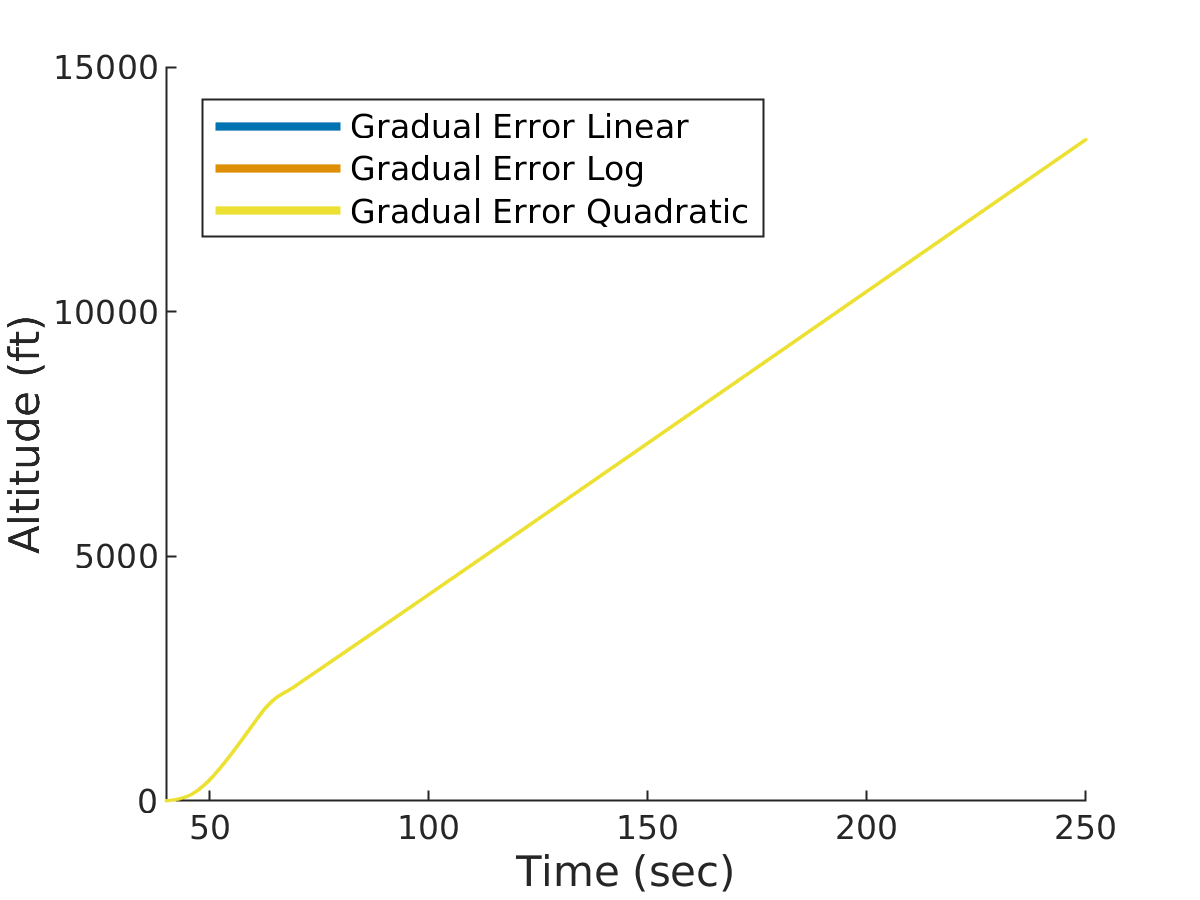}}
            \end{minipage}%
            \hfill
            \begin{minipage}[b]{0.5\columnwidth}
                \centering
                \subcaptionbox{\small Pilot-induced stall flight paths.\label{fig:sa_mcas_traj:stall}}{
                    \includegraphics[width=\textwidth]{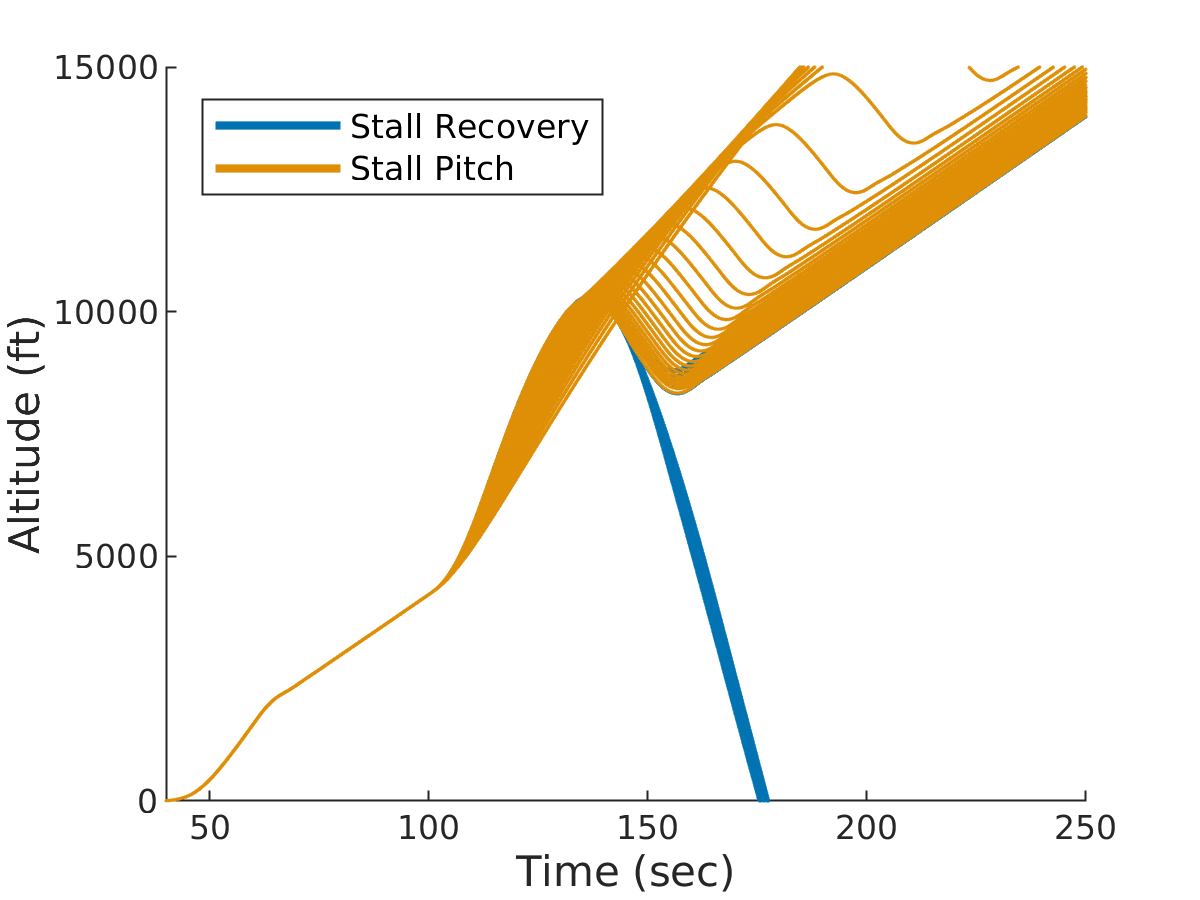}}
            \end{minipage}
        \end{minipage}%
        \hfill
        \begin{minipage}[b]{0.95\columnwidth}
            \begin{subfigure}[b]{\columnwidth}
                \centering
                \includegraphics[width=0.8\textwidth]{000-Results/RTAS-24-SAFE-STATE-OLD.pdf}
                \caption{\small Pilot can have
                some variability in their response time and
                exerted effort on the HS hand-crank. This
                figure examines the impact of this variability
                on aircraft recovery. The recoverability of the
                flight {\em is not} dictated by the pilot's reaction
                speed and rotation of the HS.}
            \label{fig:sa_mcas_safe_state}
            \end{subfigure}
            \vspace{-0.4in}
        \end{minipage}
    \caption{\small 
    Summary of the stress test simulation for \name.}
    \label{fig:sa_mcas_traj_safe_state}
    \vspace{-0.25in}
\end{figure*}

With an available simulator for streamlining the design
and evaluation of MCAS programs (\autoref{sec:sim}) and
well-defined failure scenarios (\autoref{sec:threats}),
we must consider a solution that merges the strengths
of both \mcasold{} and \mcasnew. Unlike the prior versions
of MCAS, our solution, \name, does not incorporate a
static assignment of control authority during control conflicts.
Here, we evaluate our implementation of \name.
By doing so, we seek a conclusive answer to
\textsc{Challenge}$-$\ding{184}.
The summary of the results for the stress test
of \name{} may be found in
\autoref{fig:sa_mcas_traj_safe_state}
and the third column of
\autoref{tbl:mcas_summary}.
\name{} is implemented using \autoref{alg:sa_mcas},
which was detailed in \autoref{sec:design}.




\subsection{Sensor Errors.}
For sensor errors of any type
(i.e., {\em sudden}, {\em delta}, or
{\em gradual errors}), \name{} provides the
correct control of the aircraft's HS in
{\em every single modeled sensor measurement
error scenario}
that we outlined in \autoref{sec:threats:methodology}.
Unlike \mcasold, \name{} is capable of preventing
sensor measurement errors from causing dangerous MCAS
control.
Furthermore, because \autoref{alg:sa_mcas} has two
layers of internal and external redundancy, \name{}
can prevent erroneous activation when
both AoA sensors have measurement error. This was
not possible with either \mcasold{} or \mcasnew.

\subsection{Dangerous Pilot Behavior.}
\label{sec:prevention:result:pilot}
To evaluate \name{} on dangerous pilot behavior,
we similarly use our simulated pilot maneuvers from
\autoref{sec:threats:methodology}. Comparing
the results in \autoref{fig:sa_mcas_safe_state} 
to \mcasnew, \name{} does not reduce the
amount of time the pilot has to respond in order
to successfully recover the aircraft. Here, the
results mirror the strengths of \mcasold.
These simulations demonstrate that \name's ability
of utilizing the strengths of the prior MCAS
versions, and hence enabling an overall safer
autonomous control system.

\subsection{Deadline \& Timing Analysis.}
\label{sec:prevention:result:timing}
Reverting the decision to only allow MCAS to activate
once, \name{} returns to the stall-prevention capabilities
of \mcasold. Thus, the pilot has nearly 2s more time than \mcasnew{}
to react to a stall event with \name, a $\sim$44\% increase. 
The FAA guidelines say that the pilot should react to 
a major failure within 3s, but their flight training 
materials say that it is likely for a pilot to take 
even longer to respond.
For instance, in the crashes of JT610 and ET302
the pilots were distracted and could not respond
within 3s. The benefit of providing the pilot with
more time to respond cannot be understated: in realistic 
scenarios for a safety-critical system it can be
between life and death.


\begin{tcolorbox}[width=\columnwidth, colback=black!10, arc=3mm]
    \underline{Conclusion for 
    \textsc{Challenge}$-$\ding{184}}:
        \textit{We present the evaluation of \name, an MCAS 
        that is capable of resolving control conflicts between 
        the manual and automatic input.
        It is less susceptible to the previously identified 
        control threats, increasing the upper bounds on the 
        conditions for aircraft recoverability.
        }
\end{tcolorbox}


\section{Discussion}
\label{sec:disc}




Below we discuss limitations of our work and
potential directions for addressing them.
For the cases where iterative improvements to \name{}
may be available, we leave these as future work.

{\bf Prevention of Dangerous Pilot Input is Effective in
    a Limited Scope of Conditions.}
As we alluded to in \autoref{sec:prevention:result:pilot},
there is a limited scope of conditions where \name{}
is incapable of recovering an aircraft from dangerous input.
Generally, these simulations involve a pilot reaction time
greater than 6s. In order to overcome this limitation,
MCAS would require greater control authority, which the
FAA would need to first approve. Given the constraint
of how MCAS is currently allowed to operate, 
such a drawback may be acceptable.
As mentioned in \autoref{sec:prevention:result:timing}, 
\name{} improves the state-of-the-art by improving the maximum
pilot reaction time by nearly 2s while avoiding sensor
failures and being capable of handling multiple
sensors failing at once.

{\bf Passenger Trust.}
In the aftermath of the Boeing 737-MAX crashes, passenger
trust towards the 737-MAX aircraft has slowly recovered.
The main contributor to this revival of trust has been from
dropping MCAS as a tool that is capable of having authority
to autonomously control the pitch of the aircraft.
However, Boeing continues fall under scrutiny due to
arising safety concerns for the 737-MAX.
One drawback of our work is its assumption of regaining
the trust of such passengers with a dynamic
authority \name. However, we note that the airline industry is
not the only one facing this challenge---the autonomous vehicle
industry has faced several controversies due to issues in
the self-driving algorithms that lead to deadly 
accidents.

While restoring public trust in autonomous systems is
outside the scope of this paper, we acknowledge the drawback
that this issue presents to \name. In order to regain this
trust, we propose that a system
such as \name{} should be introduced in such a way that would
(1) educate the pilot on its autonomous functions and limitations 
so the pilot will not over-trust MCAS, and
(2) give the pilot the capability to disable
\name{} in the event that issues with the algorithm arise.
Before ever being put into the air, \name{} should also go
through hundreds of simulated flight hours with real pilots
in order to establish trust with regulation agencies such
as the FAA.
While this may be seen as contradictory
to our original motivation, it may still be a
necessary measure. In fact, our argument is about the
controller capabilities and conflict, e.g., with respect
to MCAS itself. If the pilot chooses to disable
MCAS altogether, then this is outside the scope of our goals.

\section{Conclusion}
\label{sec:conc}
In this paper, we introduced \name{}, a system
for deciding who to trust when a human pilot and the
autonomous MCAS module of the Boeing 737-MAX are in
disagreement. Our analysis of the control risks of
\mcasold{} and \mcasnew{} shows the need
for an MCAS that can arbitrate control.
We demonstrate \name's capability of providing
the correct control input in all cases of injected
erroneous sensor values as well as many instances of
dangerous pilot behavior, matching the best performance
of both \mcasold{} and \mcasnew.
Our results suggest that it would be beneficial for the
flight control computer of a Boeing 737-MAX to
include a system like \name, serving as an
integrity checker that achieves the FAA's
flight directive~\cite{faa:20:faa_mcas_fix}.

\bibliographystyle{IEEEtran}
\bibliography{_ieee-format}

\begin{thebibliography}{10}
\providecommand{\url}[1]{#1}
\csname url@samestyle\endcsname
\providecommand{\newblock}{\relax}
\providecommand{\bibinfo}[2]{#2}
\providecommand{\BIBentrySTDinterwordspacing}{\spaceskip=0pt\relax}
\providecommand{\BIBentryALTinterwordstretchfactor}{4}
\providecommand{\BIBentryALTinterwordspacing}{\spaceskip=\fontdimen2\font plus
\BIBentryALTinterwordstretchfactor\fontdimen3\font minus \fontdimen4\font\relax}
\providecommand{\BIBforeignlanguage}[2]{{%
\expandafter\ifx\csname l@#1\endcsname\relax
\typeout{** WARNING: IEEEtran.bst: No hyphenation pattern has been}%
\typeout{** loaded for the language `#1'. Using the pattern for}%
\typeout{** the default language instead.}%
\else
\language=\csname l@#1\endcsname
\fi
#2}}
\providecommand{\BIBdecl}{\relax}
\BIBdecl

\bibitem{faa:19:grounding}
\BIBentryALTinterwordspacing
{Federal Aviation Administration}. (2019) {Emergency Order of Prohibition}. [Online]. Available: \url{https://web.archive.org/web/20230417205843/https://www.faa.gov/news/updates/media/Emergency_Order.pdf}
\BIBentrySTDinterwordspacing

\bibitem{boeing:19:mcas_updates}
\BIBentryALTinterwordspacing
{Boeing}. (2019) {737 MAX Software Update}. [Online]. Available: \url{https://www.boeing.com/commercial/737max/737-max-software-updates.page}
\BIBentrySTDinterwordspacing

\bibitem{gates:20:st}
D.~Gates and M.~Baker, ``{The Inside Story of MCAS: How Boeing’s 737 MAX System Gained Power and Lost Safeguards},'' \emph{The Seattle Times}, 2019.

\bibitem{lemme:21:mid_value}
\BIBentryALTinterwordspacing
P.~Lemme. (2021) {Mid-Value Select (MVS): Goldilocks in the House of MCAS}. [Online]. Available: \url{https://www.satcom.guru/2021/01/mid-value-select-mvs-goldilocks-in.html}
\BIBentrySTDinterwordspacing

\bibitem{chen:22:thesis}
C.-Y. Chen, ``{Context-Aware Detection and Resolution of Data Anomalies for Semi-Autonomous Cyber-Physical Systems},'' Ph.D. dissertation, University of Michigan, 2022.

\bibitem{xue:22:usenixsec}
L.~Xue, Y.~Liu, T.~Li, K.~Zhao, J.~Li, L.~Yu, X.~Luo, Y.~Zhou, and G.~Gu, ``{SAID: State-Aware Defense Against Injection Attacks on In-Vehicle Network},'' in \emph{USENIX Security Symposium}, 2022.

\bibitem{fei:19:tvt}
F.~Guo, Z.~Wang, S.~Du, H.~Li, H.~Zhu, Q.~Pei, Z.~Cao, and J.~Zhao, ``{Detecting Vehicle Anomaly in the Edge via Sensor Consistency and Frequency Characteristic},'' \emph{IEEE Transactions on Vehicular Technology}, vol.~68, no.~6, 2019.

\bibitem{curran:23:cns}
N.~T. Curran, A.~Ganesan, M.~D. Pesé, and K.~G. Shin, ``{Using Phone Sensors to Augment Vehicle Reliability},'' in \emph{IEEE Conference on Communications and Network Security (CNS)}, 2023.

\bibitem{ganesan:17:sae}
A.~Ganesan, J.~Rao, and K.~G. Shin, ``{Exploiting Consistency Among Heterogeneous Sensors for Vehicle Anomaly Detection},'' SAE Technical Paper, Tech. Rep., 2017.

\bibitem{checkoway:10:oakland}
{K. Koscher, S. Checkoway, et al.}, ``{Experimental Analysis of a Modern Automobile},'' in \emph{IEEE Symposium on Security \& Privacy (S\&P)}, 2010.

\bibitem{miller:15:blackhat}
C.~Miller and C.~Valasek, ``{Remote Exploitation of an Unaltered Passenger Vehicle},'' in \emph{Black Hat USA}, 2015.

\bibitem{berndt:04:aiaa}
J.~Berndt, ``{JSBSim: An Open Source Flight Dynamics Model in C++},'' in \emph{AIAA Modeling and Simulation Technologies Conference and Exhibit}, 2004.

\bibitem{boeing:operations}
{Boeing}, ``{Boeing 737-600/-700/-800/-900 Operations Manual},'' 2018.

\bibitem{how:13:mitocw}
J.~P. How, ``{Lecture notes in 16.333: Aircraft Stability and Control},'' 2004.

\bibitem{ARP4761}
{Aerospace Recommended Practice}, ``{ARP 4761: Guidelines and Methods for Conducting the Safety Assessment Process on Civil Airborne Systems and Equipment},'' SAE International, Standard, Dec. 1996.

\bibitem{ARP4754}
------, ``{ARP 4754: Guidelines For Development Of Civil Aircraft and Systems},'' SAE International, Standard, Dec. 2010.

\bibitem{jt610:18:report}
\BIBentryALTinterwordspacing
S.~Tjahjono, ``{Preliminary KNKT.18.10.35.04 Aircraft Accident Investigation Report},'' National Transportation Safety Committee of Indonesia, Tech. Rep., 2018. [Online]. Available: \url{https://downloads.regulations.gov/FAA-2024-0159-0002/attachment_11.pdf}
\BIBentrySTDinterwordspacing

\bibitem{faa:20:faa_mcas_fix}
\BIBentryALTinterwordspacing
{Federal Aviation Administration}, ``{Airworthiness Directives; The Boeing Company Airplanes},'' \emph{Federal Register}, vol.~85, no. 225, pp. 74\,560--74\,593, 2020. [Online]. Available: \url{https://www.federalregister.gov/d/2020-25844}
\BIBentrySTDinterwordspacing

\bibitem{microchip:24:tpm}
\BIBentryALTinterwordspacing
M.~Technology. (2024) {Trusted Platform Module: Complete Security for PCs and Embedded Systems}. [Online]. Available: \url{https://www.microchip.com/en-us/products/security/security-ics/tpm}
\BIBentrySTDinterwordspacing

\bibitem{jeannin:15:emsoft}
J.-B. Jeannin, K.~Ghorbal, Y.~Kouskoulas, R.~Gardner, A.~Schmidt, E.~Zawadzki, and A.~Platzer, ``{Formal verification of ACAS X, an industrial airborne collision avoidance system},'' in \emph{International Conference on Embedded Software (EMSOFT)}, 2015.

\bibitem{faa:21:flying_handbook_17}
\BIBentryALTinterwordspacing
{Federal Aviation Administration}, ``{Airplane Flying Handbook Chapter 17},'' 2021. [Online]. Available: \url{https://www.faa.gov/regulations_policies/handbooks_manuals/aviation/airplane_handbook/media/19_afh_ch17.pdf}
\BIBentrySTDinterwordspacing

\bibitem{stackexchange:19:trim_wheel}
\BIBentryALTinterwordspacing
Bianfable. (2019) {How many turns of a 737 trim wheel equal 2.5 degrees stabilizer deflection?} [Online]. Available: \url{https://aviation.stackexchange.com/questions/70184/how-many-turns-of-a-737-trim-wheel-equal-2-5-degrees-stabilizer-deflection}
\BIBentrySTDinterwordspacing

\bibitem{nasa:23:climb}
\BIBentryALTinterwordspacing
{NASA}. (2023) {Forces in a Climb}. [Online]. Available: \url{https://www1.grc.nasa.gov/beginners-guide-to-aeronautics/forces-in-a-climb/}
\BIBentrySTDinterwordspacing

\bibitem{wiki:24:drag_physics_falling_object}
\BIBentryALTinterwordspacing
{Wikipedia}. (2024) {Drag (physics): Velocity of a falling object}. [Online]. Available: \url{https://en.wikipedia.org/wiki/Drag_(physics)#Velocity_of_a_falling_object}
\BIBentrySTDinterwordspacing

\bibitem{dodt:isasi:11}
T.~Dodt, ``{Introducing the 787: Effect on Major Investigations and Interesting Tidbits},'' 2011.

\bibitem{sun:20:thesis}
K.~Sun, ``{Reliable Air Data Solutions For Small Unmanned Aircraft Systems},'' Ph.D. dissertation, University of Minnesota, 2020.

\bibitem{uscongress:20:737maxreport}
\BIBentryALTinterwordspacing
P.~A. Defazio and R.~Larson. (2020) {Final Committee Report: The Design, Development, and Certification of the Boeing 737 MAX}. [Online]. Available: \url{https://web.archive.org/web/20221207103614/https://transportation.house.gov/imo/media/doc/2020.09.15%20FINAL%20737%20MAX%20Report%20for%20Public%20Release.pdf}
\BIBentrySTDinterwordspacing

\bibitem{lie:13:jaircraft}
F.~A.~P. Lie and D.~Gebre-Egziabher, ``{Synthetic Air Data System},'' \emph{Journal of Aircraft}, vol.~50, no.~4, pp. 1234--1249, 2013.

\bibitem{klein:06:aiaa}
V.~Klein and E.~A. Morelli, ``{Aircraft System Identification: Theory and Practice},'' \emph{AIAA Education Series}, vol. 213, 2006.

\bibitem{zeis:88:thesis}
J.~E. Zeis, ``{Angle of Attack and Slideslip Estimation Using an Inertial Reference Platform},'' Master's thesis, Airforce Institute of Technology, 1988.

\bibitem{yang:23:cdc}
S.~Yang, G.~J. Pappas, R.~Mangharam, and L.~Lindemann, ``{Safe Perception-Based Control under Stochastic Sensor Uncertainty using Conformal Prediction},'' in \emph{IEEE Conference on Decision and Control (CDC)}, 2023.

\bibitem{jackson:15:aiaa}
{E. B. Jackson, et al.}, ``{Further Development of Verification Check-Cases for Six-Degree-of-Freedom Flight Vehicle Simulations},'' in \emph{AIAA Modeling and Simulation Technologies Conference}, 2015.

\bibitem{sikstom:21:thesis}
T.~Sikström, ``{Flight Simulator Integration in Test Rig},'' Master's thesis, Royal Institute of Technology, 2021.

\bibitem{krepelka:22:b737800_takeoff}
\BIBentryALTinterwordspacing
M.~Křepelka. (2022) {Boeing 737–800 Flight Notes}. [Online]. Available: \url{https://krepelka.com/fsweb/learningcenter/aircraft/flightnotesboeing737-800.htm}
\BIBentrySTDinterwordspacing

\bibitem{pprn:10}
\BIBentryALTinterwordspacing
{Badmachine}. (2010) {Built-In Bank Angle Limits}. [Online]. Available: \url{https://www.pprune.org/tech-log/416502-built-bank-angle-limits.html}
\BIBentrySTDinterwordspacing

\bibitem{faa:22:holding_rules}
\BIBentryALTinterwordspacing
{Federal Aviation Administration}. (2022) {ENR 1.5 Holding, Approach, and Departure Procedures}. [Online]. Available: \url{https://www.faa.gov/air_traffic/publications/atpubs/aip_html/part2_enr_section_1.5.html}
\BIBentrySTDinterwordspacing

\bibitem{williams:21:b737800_takeoff}
\BIBentryALTinterwordspacing
I.~D. Williams. (2021) {Boeing 737-800 Takeoff Procedure (simplified)}. [Online]. Available: \url{https://www.flaps2approach.com/journal/2014/8/4/boeing-737-800-takeoff-procedure-simplified.html}
\BIBentrySTDinterwordspacing

\bibitem{jt610:18}
\BIBentryALTinterwordspacing
{National Transportation Safety Committee of Indonesia}, ``{Accident Boeing 737 MAX 8, PK-LQP (LNI610)},'' 2018. [Online]. Available: \url{https://ngamotu.nz/images/20181122-jt610-knkt.pdf}
\BIBentrySTDinterwordspacing

\bibitem{jt610:18:flightradar}
\BIBentryALTinterwordspacing
I.~Petchenik. (2018) {JT610 Granular ADS-B Data}. Flightradar24. [Online]. Available: \url{https://www.flightradar24.com/blog/wp-content/uploads/2018/10/JT610_Granular_ADSB_Data.csv}
\BIBentrySTDinterwordspacing

\end{thebibliography}


\end{document}